%% file: paper_v2.tex
\documentclass[12pt,psfig]{article}
\usepackage{a4wide}
\usepackage{graphicx}
\usepackage{subfigure}
\usepackage[centertags]{amsmath}
\usepackage{amssymb}
\usepackage{dsfont}
\newcommand{\unity}{{ 1\:\!\!\!\mbox{I}}}

\newcommand{\Z}{{\mathbb Z}}

\newcommand{\OR}{\Omega {\mathcal R}}

\newcommand{\ku}{\mathcal{K}}
\newcommand{\au}{\mathcal{A}}
\newcommand{\msu}{\mathcal{M}}

\newcommand{\SC}{\scriptstyle}

\newcommand{\ov}{\overline}

\newcommand{\barre}[1]{%
        \setbox1=\hbox{$#1$} \dimen2=\ht1 \dimen3=\dp1 \dimen4=\wd1
        \setbox2=\hbox{\sl /}
        \dimen1=\wd1 \advance\dimen1 by -\wd2 \divide\dimen1 by 2
        \advance\dimen1 by \wd2 \advance\dimen1 by 0.4pt
        \setbox3=\hbox to \wd1{\hss \box1 \kern -\dimen1 \box2\hss}
        \ht3=\dimen2 \dp3=\dimen3 \wd3=\dimen4
        \box3
        }
\begin{document}
\pagestyle{empty}
\rightline{FT-UAM-03-04}
\rightline{IFT-UAM/CSIC-03-08}
\begin{flushright}
                     hep-th/0303015
\end{flushright}
\vskip 2cm

\begin{center}
{\huge Chiral supersymmetric models on\\}
{\huge an orientifold of $\Z_4 \times \Z_2$\\}
{\huge with intersecting D6-branes}
\vspace*{5mm} \vspace*{1cm}
\end{center}
\vspace*{5mm} \noindent
\vskip 0.5cm
\centerline{\bf Gabriele Honecker}
\vskip 1cm
\centerline{\em Departamento de F\'{\i}sica Te\'{o}rica C-XI and 
Instituto de F\'{\i}sica Te\'{o}rica C-XVI,}
\centerline{\em Universidad Aut\'{o}noma de Madrid, Cantoblanco, 28049 Madrid, Spain} 
\vskip2cm

\centerline{\bf Abstract}

We investigate 
intersecting D6-branes on an orientifold of type IIA theory in the orbifold background $T^6/(\Z_4 \times \Z_2)$
with the emphasis on finding chiral spectra. RR tadpole cancellation conditions and the 
scalar potential at disc level are computed. The general 
chiral spectrum is displayed, and two supersymmetric models with a Pati-Salam group are shown, one with four generations and the 
other one with three generations and {\it exactly} the chiral matter content of the SM plus right handed neutrinos.

\vskip .3cm

%%%%%%%%%%%%%%%%%%%%%%%%%%%%%%%%%%%%%%%%%%%%%%%%%%%%%%%%%%%%%%%%%%%%%%%%%%%%%

\newpage

\setcounter{page}{1} \pagestyle{plain}

\section{Introduction} \label{intro}

In the recent years, open string models have started to play a key role in constructing 
phenomenologically appealing string vacua as well as getting a deeper understanding of the still
unknown M-theory. The two major approaches in this direction consist in constructing chiral theories
from D-brane set-ups in type II orientifold theories. One possibility relies on D-branes located at 
singularities which support chiral fermions on their worldvolume, see e.g.~\cite{Angelantonj:1996uy,Klein:2000qw}.
In the other approach, D-branes wrap some compact dimensions, and chiral fermions are located at 
intersections of D-branes~\cite{Berkooz:1996km} wrapping different compact cycles. 
The search for chiral and stable vacua in this framework has turned out to
be be highly non-trivial. The first models with  D-branes intersecting at angles were those where 
the D-branes are located on top of orientifold planes also at angles in an orbifold background. All these models
turned out to be supersymmetric but non-chiral~\cite{Blumenhagen:1999md,Forste:2000hx}.
Subsequently, general features for model building within intersecting
brane world scenarios, e.g. family replication as well as the leading
order behavior of gauge and Yukawa couplings in terms of geometric
quantities of the compactification, were 
worked out in~\cite{Blumenhagen:2000fp,Blumenhagen:2000ea,Aldazabal:2000cn}. 

The picture of D-branes at non-trivial angles has a T-dual formulation in terms of D-branes with 
different magnetic background fluxes. An early observation on the 
possibility of supersymmetry breaking in this language was made in~\cite{Bachas:1995ik}. Further works 
in the picture with background fluxes were performed in~\cite{Angelantonj:2000hi}.

Pioneered by~\cite{Ibanez:2001nd}, searches for exact realizations of the standard model and 
GUT models from intersecting D6-branes were 
performed in a number of papers~\cite{Blumenhagen:2001te,Kokorelis:2002ip}, 
and phenomenological issues for these models, e.g. the appearance of
additional $U(1)$ symmetries and the interpretation of the Higgs
mechanism as brane
recombination,  were addressed in more detail~\cite{Cremades:2002te}; 
for a very recent discussion on the exact structure of Yukawa couplings see also~\cite{Cremades:2003qj}.

Motivated by large extra dimension scenarios~\cite{Arkani-Hamed:1998rs},
models with intersecting D4- and D5-branes with and without orientifold projections in the type II string
theory framework were as well investigated in~\cite{Bailin:2001ie}.
A related model within the type 0' theory was discussed in~\cite{Blumenhagen:2002mf}, and applications to
cosmology started to be considered (see e.g.~\cite{Garcia-Bellido:2001ky}).
 
In~\cite{Blumenhagen:2002wn}, generalizations of intersecting D-brane set-ups beyond the 
conformal field theory limit on orbifolds were discussed.

In general, the chiral models are non-supersymmetric. Even if the low energy spectra are free of tachyons, 
NSNS tadpoles signal that the analysis is performed in a false vacuum. Furthermore, 
D6-branes at angles do not allow 
for large extra dimensions but enforce the string scale to be of the order of the Planck scale.

These problems do, however, not apply to supersymmetric models. In~\cite{Cvetic:2001tj}, 
a class of chiral supersymmetric models with intersecting D6-branes in the orbifold background
$T^6/(\Z_2 \times \Z_2)$ was constructed. The phenomenology for standard model like spectra was subsequently 
worked
out in~\cite{Cvetic:2002qa}. A systematic search for $SU(5)$ GUT models has also been 
performed recently~\cite{Cvetic:2002pj}.

Together with the $T^6/\Z_4$ orientifold model in~\cite{Blumenhagen:2002gw}, these are the two only known
cases of models with D6-branes intersecting at angles which preserve supersymmetry and provide 
phenomenologically appealing chiral spectra but, however, contain some
exotic matter.

Last but not least, another motivation to proceed with the search for chiral supersymmetric 
vacua from intersecting D6-branes consists in the fact that only the supersymmetric case 
lifts to a purely geometrical background in M-theory on a manifold with $G_2$ holonomy
(see e.g.~\cite{Uranga:2002ag} and the corresponding sections in~\cite{Cvetic:2001tj,Blumenhagen:2002wn}
plus references therein).

In this paper, we search for chiral supersymmetric models in a type IIA  orientifold on $T^6/(\Z_4 \times \Z_2)$.
The most simple non-chiral model on this orbifold was already found in~\cite{Forste:2000hx}.
Many features are similar to the simpler chiral model with product orbifold group 
$T^6/(\Z_2 \times \Z_2)$~\cite{Cvetic:2001tj}, but as
in the case of $T^6/\Z_4$~\cite{Blumenhagen:2002gw}, the D6-branes lie in certain invariant orbits under the
orbifold group which make the analysis more involved.

The paper is organized as follows: in section~\ref{geometry}, the geometry of the compactification 
on $T^6/(\Z_4 \times \Z_2)$ and 
basic model building ingredients are described.
In section~\ref{RRtadpoles}, details of the RR tadpole cancellation calculation 
and the required D6-brane configuration 
are given.
Subsequently, in section~\ref{chiralspectrum} the chiral spectrum is displayed. NSNS tadpoles are briefly
commented on in section~\ref{NSNStadpoles}. Two chiral models with a
Pati - Salam gauge group --- including one with three generations and
no exotic chiral matter --- are
discussed in section~\ref{SUSYmodels}. Finally, the conclusions are given in section~\ref{Conclusions}
and some further technical details of the calculation are collected in two appendices~\ref{AppIntersectionNumbers}
and~\ref{AppSusycond}.

\section{Geometry of the $T^6/(\OR \times \Z_4 \times \Z_2)$ orientifold} \label{geometry}

Throughout the paper, we consider an orientifold of type IIA string theory compactified 
on the orbifold $T^6/ (\Z_4 \times \Z_2)$.
The orbifold group $\Z_4\times \Z_2$ is generated by
\begin{equation}
\begin{aligned} 
\Theta: \qquad &(z^1,z^2,z^3) \rightarrow (iz^1,-iz^2,z^3),\\  
\omega: \qquad &(z^1,z^2,z^3) \rightarrow (z^1,-z^2,-z^3),
\end{aligned}\nonumber
\end{equation}
where $z^1=x^4+ix^5$, $z^2=x^6+ix^7$, $z^3=x^8+ix^9$ label the internal complex coordinates
 of the six-torus which we consider to be decomposed into a  product of three two-tori, 
 \mbox{$T^6 \simeq T^2_1 \times T^2_2 \times T^2_3$.}
This orbifold group action describes the singular limit of a Calabi-Yau compactification
preserving ${\cal N}=2$ supersymmetry in four dimensions. The Hodge numbers of this manifold are 
$h_{1,1}=61$ and $h_{2,1}=1$.\footnote{I am grateful to K.~Wendland for the computation of 
the Hodge numbers. See also~\cite{Klein:2000qw}.} 
The number of independent 3-cycles is given by $b_3=2+2h_{2,1}=4$, which means that
all 3-cycles of this model are inherited from the six-torus $T^6$. This is in contrast to the $T^6/\Z_4$ orbifold
which was considered in~\cite{Blumenhagen:2002gw} where also exceptional cycles have to be included in 
the analysis of the underlying geometry. Furthermore, three $(1,1)$-forms are inherited from the
 torus, and $18+12+12$ arise from fixtori under $\Theta^2, \omega$ and $\Theta^2\omega$, respectively. 
The fixedpoints of $\Theta\omega$ provide 16 further $(1,1)$-forms.

In addition to the orbifold group, we perform an orientifold projection $\OR$ which was first employed 
in~\cite{Blumenhagen:1999md,Forste:2000hx}. The worldsheet 
parity $\Omega$ is accompanied by a geometric action on the compact space,
\begin{equation}
{\cal R}: \qquad z^i \longrightarrow z^{\ov{i}}, \qquad i=1,2,3.     \nonumber
\end{equation}
This projection introduces O6-planes which are located at the 3-cycles invariant under 
${\cal R}\Theta^k\omega^l$ for $k=0,\ldots,3$ and $l=0,1$. In order to cancel the RR tadpoles which arise 
from the O6-planes, one is forced to include D6-branes intersecting at angles in the model. 
An explicit NSNS three-form flux $H_{NS}$ might also play a role in canceling chiral anomalies~\cite{Uranga:2002vk},
which, however, is not taken into account in this paper since the search here is focused on 
supersymmetric models.
In the most simple example of this 
kind, the D6-branes lie on top of the O6-planes and cancel all RR-charges locally leading to a purely 
non-chiral spectrum as investigated in~\cite{Forste:2000hx}.

\begin{figure}
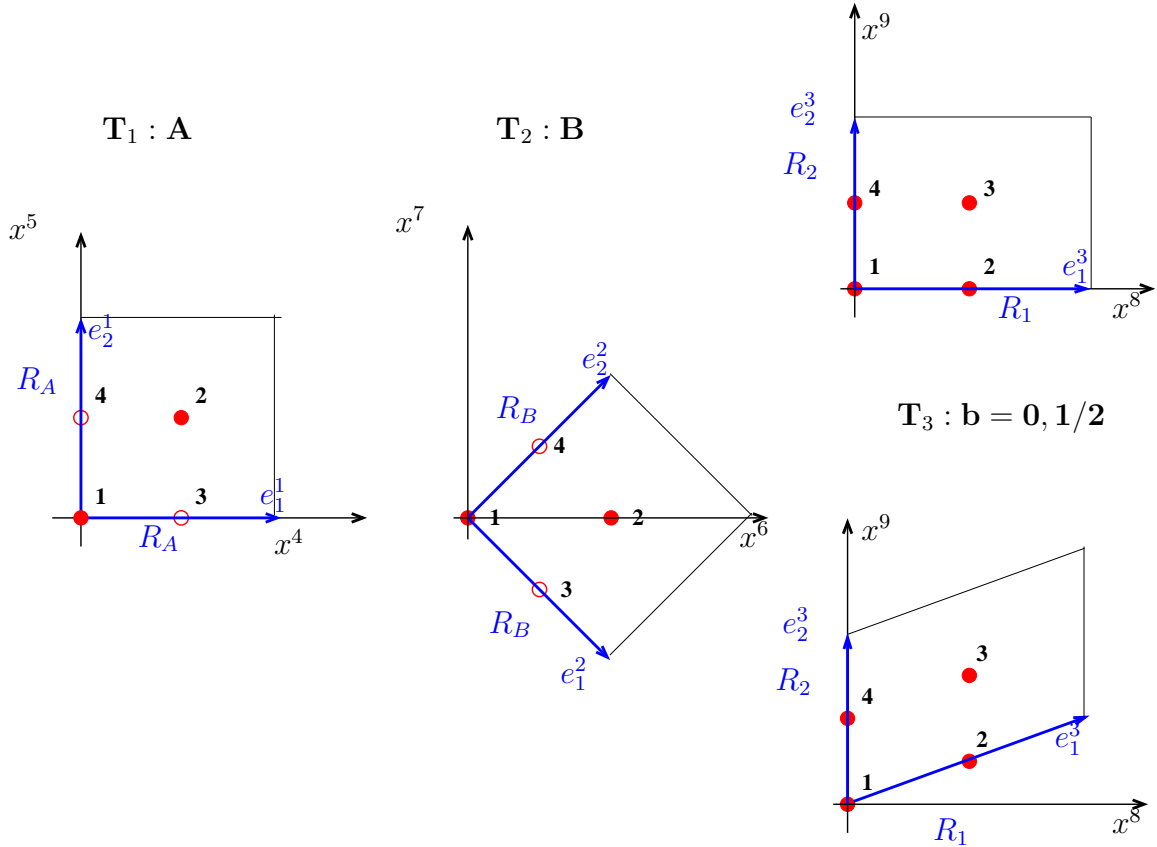

\begin{center}
\input Z4xZ2fixedpoints.pstex_t
\end{center}
\caption{The possible choices of the compactification lattices per two-torus. $\Z_4$ fixed points are
depicted by filled circles, the additional points invariant under $\Z_2$ subsymmetries by empty circles. 
The basis of each two-torus in our convention is displayed in blue. The two possible choices of the third
lattice correspond to a vanishing and non-trivial antisymmetric tensor background
on $\Tilde{T}^2_3$ in the T-dual type IIB orientifold, respectively.}
\label{FigureZ4xZ2fixedpoints}
\end{figure}

The constraints on the shape of the two-tori imposed by the orbifold group $\Z_4 \times \Z_2$ and the
anti-holomorphic involution ${\cal R}$ are the same as for the $\Z_4$ model considered 
in~\cite{Blumenhagen:2002gw}.
Namely, the complex structure moduli on the first two tori $T^2_1\times T^2_2$ are fixed by the $\Z_4$ 
symmetry such as to parameterize square tori. Furthermore, the reflexion  ${\cal R}$ enforces the orientation 
of the lattices to be either of type {\bf A} or {\bf B} as depicted in figure~\ref{FigureZ4xZ2fixedpoints}.
The radii $R_A$ and $R_B$ remain as moduli of the compactification. 
The third torus $T^2_3$ is not constrained by the orbifold action since
it is only subject to a $\Z_2$ rotation. The reflexion  ${\cal R}$, however, constrains its shape. The two
allowed choices correspond in the T-dual type IIB orientifold model, where T-duality is taken to act along the $x^9$ 
coordinate,
to the possibility of having either a vanishing or a non-trivial quantized antisymmetric tensor background 
on the dual torus $\Tilde{T}^2_3$ conveniently parameterized by $b=0,1/2$, respectively~\cite{Blumenhagen:2000ea}. 
In addition, the imaginary part of the complex structure and the volume 
are moduli which can be parameterized by the two radii $R_1$ and $R_2$ as depicted in 
figure~\ref{FigureZ4xZ2fixedpoints}.
%%%%%%%%%%%%%%%%%%%%%%%%%%%%%%%%%%%%%%%%%%%%%
%%%%%%%%%%%%%%%%%%%%%%%%%%%%%%%%%%%%%%%%%%%%

The factorizable 3-cycles which the O$6_a$-planes and D$6_a$-branes wrap are conveniently characterized by the 
wrapping numbers $(n^a_i,m^a_i)$  on the $i^{\SC th}$ two-torus $T^2_i$ along the basis $(e^i_1,e^i_2)$
displayed in figure~\ref{FigureZ4xZ2fixedpoints}. In general, different kinds of D6-branes $a$ and $b$
intersect several times on the fundamental cell of the torus. The intersection number 
in terms of wrapping numbers 
is given by
\begin{equation}
I_{ab}=\prod_{i=1,2,3}(n^a_i m^b_i-n^b_i m^a_i).   \nonumber
\end{equation} 
The symmetries ${\cal R}$ and $\Theta$ enforce the positions of O6-planes and D6-branes to be grouped into 
invariant orbits as described in more detail in section~\ref{RRtadpoles} in the course of calculating the 
Klein bottle and M\"obius strip amplitudes.

\section{RR tadpole cancellation} \label{RRtadpoles}

The requirement of vanishing net RR-charge constrains the allowed sets of wrapping numbers and 
numbers of identical D6-branes. The values depend on the shape of the third torus or, in the T-dual picture, 
on the value of the antisymmetric tensor background.

\subsubsection*{The Klein bottle amplitude} 

The Klein bottle amplitude can be easily obtained from the computation in~\cite{Forste:2000hx} 
where the calculation
 had been constrained to the case $R_2=4^b R_1$ which parameterizes a 
square torus  $T^2_3$ with orientation {\bf A} or {\bf B} for $b=0,1/2$, respectively.
There exist in total three inequivalent choices of complex structures on $T^2_1 \times T^2_2$, namely 
{\bf AA}, {\bf AB} and {\bf BB}.
As in~\cite{Forste:2000hx}, we only consider the case where the lattice on 
$T^2_1 \times T^2_2$ is chosen to be of the kind {\bf AB} since, only in this case, the worldsheet 
duality is well understood in terms of introducing one sort of boundary and crosscap states. 

\begin{figure}
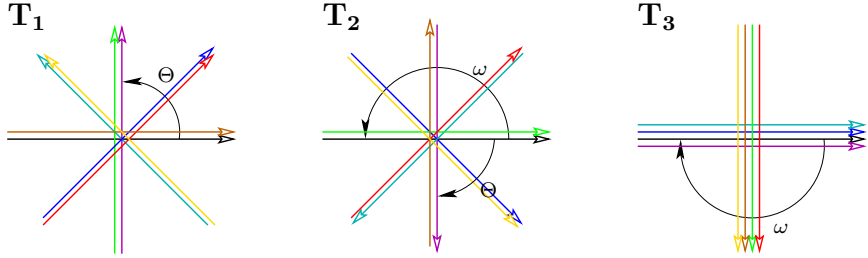

\begin{center}
\input Z4xZ2Oplanes.pstex_t
\end{center}
\caption{The four orbits of O6-planes. Each orbit consists of an O6-plane and its image under the 
$\Z_4$ generator $\Theta$. The black and the purple cycles belong to one orbit, the 
red and the yellow cycles form another orbit. The remaining orbits consist of the blue and cyan cycles and the 
green and brown ones.}
\label{FigureO6planes}
\end{figure}

The eight kinds of O6-planes required by the $\OR \times \Z_4 \times \Z_2$ symmetry are displayed in 
figure~\ref{FigureO6planes}. They are grouped into four orbits of two kinds of O6-planes each which are 
mapped onto each other under the $\Z_4$ generator $\Theta$.
The O6-planes wrapping the $x^8$ direction contribute a constant times $R_1/R_2$ to the tadpoles due
to the windings perpendicular and momenta parallel to its position. In the same spirit, the contribution originating
from the O6-planes wrapping the $x^9$ direction are proportional to $R_2/R_1$.

%%%%%%%%%%%%%%%%%%%%%%%%%%%%%%%%%%%%%%%%%%%%%
%%%%%%%%%%%%%%%%%%%%%%%%%%%%%%%%%%%%%%%%%%%%
The resulting tadpole of the Klein bottle amplitude in the tree channel is given by
\begin{equation}
\ku \rightarrow (1_{RR}-1_{NSNS}) 2^9 c
\left(\frac{R_1}{R_2}+\frac{1}{16^b}\frac{R_2}{R_1}\right)
\int_0^{\infty}  dl.\label{FormulaRRtadpoleKb}
\end{equation} 

\subsubsection*{The M\"obius strip amplitude}

\begin{figure}
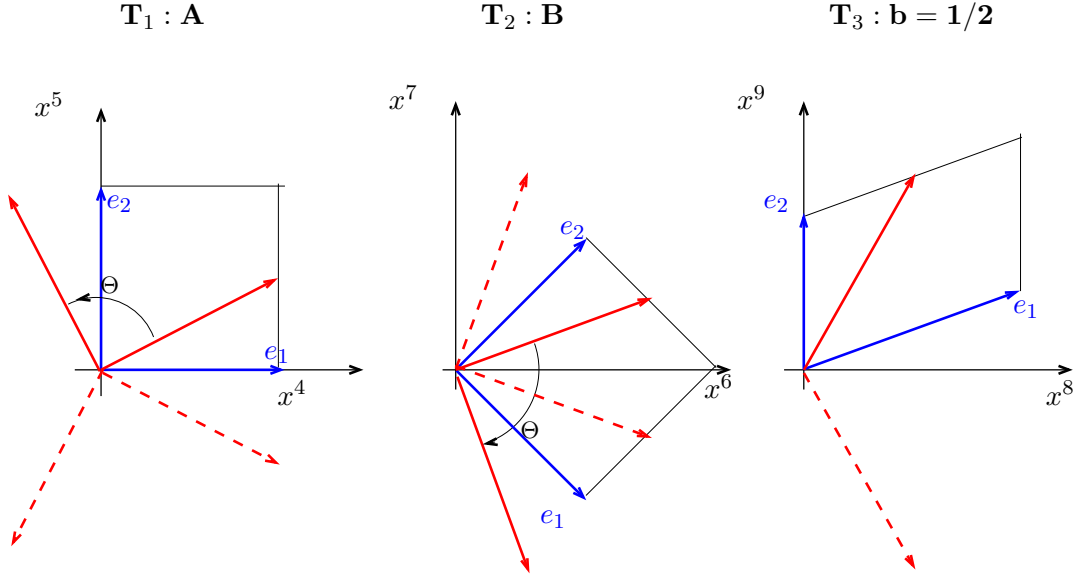

\begin{center}
\input Z4xZ2braneimages.pstex_t
\end{center}
\caption{Orbit of a D6-brane with wrapping numbers $(n^a_1,m^a_1)=(2,1)$, $(n^a_2,m^a_2)=(1,2)$, 
$(n^a_3,m^a_3)=(1,2)$ on a tilted torus $T^2_3$, i.e. $b=1/2$. The brane $a$ and its $\Z_4$ image 
$(\Theta a)$ are represented 
by solid red lines whereas the mirror images $a'$ and $(\Theta a)'$ are depicted by dashed red lines.}
\label{FigureBraneImages}
\end{figure}

The general orbit of a D6-brane consists of four different cycles. A brane $a$ is mapped under the $\Z_4$ 
rotation to its image $(\Theta a)$ and under ${\cal R}$ these two are mapped onto their mirror images 
$a'$ and $(\Theta a)'$. The wrapping numbers of such an orbit are given by 
\begin{equation}
\begin{aligned} 
a &\Leftrightarrow \left(\begin{array}{cc}
n^a_1, &m^a_1 \\ n^a_2, &m^a_2 \\   n^a_3, &m^a_3
\end{array}\right), \qquad
%%%%%%%%%%%%%%%%%%%%%%%%%%%%%%%%%%
(\Theta a)\Leftrightarrow  \left(\begin{array}{cc}
-m^a_1, &n^a_1 \\ m^a_2, &-n^a_2 \\   n^a_3, &m^a_3
\end{array}\right), \\
%%%%%%%%%%%%%%%%%%%%%%%%%%%%%%%%%%
a' &\Leftrightarrow  \left(\begin{array}{cc}
n^a_1, &-m^a_1 \\ m^a_2, &n^a_2 \\   n^a_3, &-m^a_3-2bn^a_3
\end{array}\right), \qquad
%%%%%%%%%%%%%%%%%%%%%%%%%%%%%%%%%%
(\Theta a)'\Leftrightarrow  \left(\begin{array}{cc}
-m^a_1, &-n^a_1 \\ -n^a_2, &m^a_2 \\   n^a_3, &-m^a_3-2bn^a_3
\end{array}\right).
\end{aligned} \nonumber
\end{equation}
An example of a whole D6-brane orbit is depicted in figure~\ref{FigureBraneImages}.
%%%%%%%%%%%%%%%%%%%%%%%%%%%%%%%%%%%%%%%%%%%%%
%%%%%%%%%%%%%%%%%%%%%%%%%%%%%%%%%%%%%%%%%%%%%

The M\"obius strip amplitude has contributions from two different kinds of strings: the $aa'$ strings ---
which are identified with the $(\Theta a)(\Theta a)'$ strings via a $\Z_4$ rotation ---
are invariant under $\OR\Theta^{2k}\omega^l$. The $a (\Theta a)'$ strings and their rotated images 
$(\Theta a) a'$ correspond to 
`twisted open strings' in the language of~\cite{Forste:2000hx} and are invariant under 
$\OR\Theta^{2k+1}\omega^l$.

The loop channel expression for the $aa'$ strings is given by
\begin{equation}
\begin{aligned}
%%%%   Ausdruck loop channel aa'  %%%%%%%%%%%%%%%%%%5
\msu_{aa'}=& \frac{c}{2^5}\sum_{k,l=0}^1
 i (-1)^{l+1}e^{2 i\sum_{j=1}^3 \delta^{(k,l)}_j \varphi_j}
\text{tr}\left(\gamma_{\OR\Theta^{2k}\omega^l}^{-T,a'}\gamma_{\OR\Theta^{2k}\omega^l}^{a}\right)
I^{\OR\Theta^{2k}\omega^l}_{aa'}  \\
& \times 
\int_0^{\infty}\frac{dt}{t^3}
\frac{\vartheta \Bigl[ \begin{array}{c} \SC \frac{1}{2} \\ \SC 0 \end{array} \Bigr]}{\eta^3}
\prod_{j=1}^3 
\frac{\vartheta \Bigl[\!\! \begin{array}{c}  \SC \frac{1}{2}+2\frac{\varphi_j}{\pi} \\ \SC \frac{1-\delta^{(k,l)}_j}{4}-\frac{\varphi_j}{\pi}    
\end{array} \!\!\Bigr]}
  {\vartheta \Bigl[\!\! \begin{array}{c}  \SC \frac{1}{2}+2\frac{\varphi_j}{\pi} \\  \SC \frac{1+\delta^{(k,l)}_j}{4}-\frac{\varphi_j}{\pi}
 \end{array} \!\!\Bigr]}
(t-\frac{i}{2})  \nonumber
\end{aligned}
\end{equation}
with the phases $\delta^{(k,l)}_1=(-1)^k$, $\delta^{(k,l)}_2=(-1)^{k+l}$ and $\delta^{(k,l)}_3=(-1)^l$.
$I^{\OR\Theta^{2k}\omega^l}_{aa'}$ denotes the numbers of intersections between the branes $a$ and $a'$ which 
are invariant under ${\cal R}\Theta^{2k}\omega^l$ and $\varphi_j$ is the angle of brane $a$ with respect to
the ${\cal R}$ invariant axis on the $j^{\SC th}$ two-torus $T^2_j$.
Performing the modular transformation $t=\frac{1}{8l}$ yields the tree channel expression of the 
amplitude,
\begin{equation}
%%%%   Ausdruck tree channel aa'  %%%%%%%%%%%%%%%%%%5
\msu_{aa'}=-\frac{c}{2}\sum_{k,l=0}^1 (-1)^l
\text{tr}\left(\gamma_{\OR\Theta^{2k}\omega^l}^{-T,a'}\gamma_{\OR\Theta^{2k}\omega^l}^{a}\right)
 I^{\OR\Theta^{2k}\omega^l}_{aa'}  \int_0^{\infty}dl
\frac{\vartheta \Bigl[ \begin{array}{c} \SC \frac{1}{2} \\ \SC 0 \end{array} \Bigr]}{\eta^3}
\prod_{j=1}^3 
\frac{\vartheta \Bigl[\!\! \begin{array}{c}  \SC \frac{1}{2} \\ \SC \frac{1-\delta^{(k,l)}_j}{4}+\frac{\varphi_j}{\pi}    
\end{array} \!\!\Bigr]}
  {\vartheta \Bigl[\!\! \begin{array}{c}  \SC \frac{1}{2} \\  \SC \frac{1+\delta^{(k,l)}_j}{4}+\frac{\varphi_j}{\pi}
 \end{array} \!\!\Bigr]}
(2l-\frac{i}{2}).\label{FormulaTreeMSaa}
\end{equation}
The contribution to the M\"obius strip amplitude originating from the $a(\Theta a)'$ strings is computed 
along the same lines. The loop channel expression is given by
\begin{equation}
\begin{aligned}
%%%%   Ausdruck loop channel a(\Theta a)'  %%%%%%%%%%%%%%%%%%5
\msu_{a(\Theta a)'}=  & \frac{c}{2^5}\sum_{k,l=0}^1 
 i (-1)^{l+1} e^{2 i\sum_{j=1}^3 \epsilon^{(k,l)}_j \varphi_j}I^{\OR\Theta^{2k+1}\omega^l}_{a(\Theta a)'} 
\text{tr}\left(\gamma_{\OR\Theta^{2k+1}\omega^l}^{-T,(\Theta a)'}\gamma_{\OR\Theta^{2k+1}\omega^l}^{a}\right)  \\
& \times \int_0^{\infty}\frac{dt}{t^3}
\frac{\vartheta \Bigl[ \begin{array}{c} \SC \frac{1}{2} \\ \SC 0 \end{array} \Bigr]}{\eta^3}
\left(\prod_{j=1}^2 
\frac{\vartheta \Bigl[\!\! \begin{array}{c}  \SC 2\frac{\varphi_j}{\pi} \\ \SC -\frac{\epsilon^{(k,l)}_j}{4}-\frac{\varphi_j}{\pi}   
\end{array} \!\!\Bigr]}
  {\vartheta \Bigl[\!\! \begin{array}{c}  \SC 2\frac{\varphi_j}{\pi} \\  \SC \frac{\epsilon^{(k,l)}_j}{4}-\frac{\varphi_j}{\pi}
 \end{array} \!\!\Bigr]}
\right)
\frac{\vartheta \Bigl[\!\! \begin{array}{c}  \SC \frac{1}{2}+2\frac{\varphi_3}{\pi} \\ \SC \frac{1-\epsilon^{(k,l)}_3}{4}-\frac{\varphi_3}{\pi}  
\end{array} \!\!\Bigr]}
  {\vartheta \Bigl[\!\! \begin{array}{c}  \SC \frac{1}{2}+2\frac{\varphi_3}{\pi} \\  \SC \frac{1+\epsilon^{(k,l)}_3}{4}-\frac{\varphi_3}{\pi}
 \end{array} \!\!\Bigr]}
(t-\frac{i}{2})
\end{aligned} \nonumber
\end{equation}
with the phases $\epsilon^{(k,l)}_1=-(-1)^k$, $\epsilon^{(k,l)}_2=(-1)^{k+l}$ and $\epsilon^{(k,l)}_3=(-1)^l$.
Upon modular transformation, one obtains the tree channel expression
\begin{equation}
\begin{aligned}
%%%%   Ausdruck tree channel a(\Theta a)'  %%%%%%%%%%%%%%%%%%5
\msu_{a(\Theta a)'}= - & \frac{c}{2}\sum_{k,l=0}^1 (-1)^l
\text{tr}\left(\gamma_{\OR\Theta^{2k+1}\omega^l}^{-T,(\Theta a)'}\gamma_{\OR\Theta^{2k+1}\omega^l}^{a}\right)
 I^{\OR\Theta^{2k+1}\omega^l}_{a(\Theta a)'}  \\
& \times \int_0^{\infty}dl
\frac{\vartheta \Bigl[ \begin{array}{c} \SC \frac{1}{2} \\ \SC 0 \end{array} \Bigr]}{\eta^3}
\left(\prod_{j=1}^2 
\frac{\vartheta \Bigl[\!\! \begin{array}{c}  \SC \frac{1}{2} \\ \SC \frac{\epsilon^{(k,l)}_j}{4}+\frac{\varphi_j}{\pi}    
\end{array} \!\!\Bigr]}
  {\vartheta \Bigl[\!\! \begin{array}{c}  \SC \frac{1}{2} \\  \SC -\frac{\epsilon^{(k,l)}_j}{4}+\frac{\varphi_j}{\pi}
 \end{array} \!\!\Bigr]}
\right)
\frac{\vartheta \Bigl[\!\! \begin{array}{c}  \SC \frac{1}{2} \\ \SC \frac{1-\epsilon^{(k,l)}_3}{4}+\frac{\varphi_3}{\pi}    
\end{array} \!\!\Bigr]}
  {\vartheta \Bigl[\!\! \begin{array}{c}  \SC \frac{1}{2} \\  \SC \frac{1+\epsilon^{(k,l)}_3}{4}+\frac{\varphi_3}{\pi}
 \end{array} \!\!\Bigr]}
(2l-\frac{i}{2}).
\end{aligned}\label{FormulaTreeMSaThetaa}
\end{equation}
Summing over all invariant brane configurations, (\ref{FormulaTreeMSaa}) and (\ref{FormulaTreeMSaThetaa})
lead to the following contribution of the total M\"obius strip amplitude to the RR tadpole,
\begin{equation}
%%%%%%%%   Asymptotik gesamt Moebius strip tree channel   %%%%%%%%%%%%%%%%
\msu_{RR} \rightarrow 
-2^5 c \left(\frac{R_1}{R_2}\sum_{a}N_a(n^a_1N^a_2+m^a_1M^a_2)n^a_3
+\frac{R_2}{R_1}\sum_{a}N_a(n^a_1M^a_2-m^a_1N^a_2)\frac{M^a_3}{4^b}\right)
\int_0^{\infty}dl
,\label{FormulaRRtadpoleMS}
\end{equation}
where the following abbreviations for linear combinations of wrapping numbers have been used 
in order to shorten the formulas
\begin{equation}
\begin{aligned}
N^a_2 &= n^a_2+m^a_2,\\
M^a_2 &= n^a_2-m^a_2,\\
M^a_3 &= m^a_3+bn^a_3.
\end{aligned}\label{EqWrappingAbb}
\end{equation}
$N_a=\pm\text{tr}
\left(\gamma_{\OR\Theta^{k}\omega^l}^{-T}\gamma_{\OR\Theta^{k}\omega^l}\right)$
denotes the number of identical D6-branes. The signs of the traces are chosen such that the sum of RR tadpoles 
of Klein bottle, M\"obius strip and annulus amplitude gives a sum of perfect squares. The actual 
choice of Chan Paton matrices is listed in section~\ref{chiralspectrum}, equation~(\ref{EqChanPatonMatrices}).

\subsubsection*{The annulus amplitude}

The annulus amplitude is computed in the same way as for the toroidal models studied 
in~\cite{Blumenhagen:2000fp}. 
All $\Z_2 \times \Z_2$ insertions  $\Theta^{2k}\omega^l$ in the 
loop channel preserve 
any configuration of D6-branes but  lead to $\Z_2$ twisted 
RR charges which cannot be compensated for by the Klein bottle and M\"obius strip
amplitudes. Therefore,
as in the previously studied cases with $\Z_2$ subsymmetries~\cite{Blumenhagen:1999md,Forste:2000hx,Blumenhagen:2000ea,Cvetic:2001tj}, the prefactors of the amplitudes 
with $\Theta^{2k}\omega^l$ insertions have to vanish, i.e. $\text{tr}\gamma_{\Theta^{2k}\omega^l}=0$
for all D6-branes.

Taking into account all contributions to the total annulus amplitude coming from the D6-branes as well
as their $\Theta$ and ${\cal R}$ orbits, the resulting RR tadpole is given by 
\begin{equation}
%%%%%%%%   Asymptotik gesamt Annulus tree channel   %%%%%%%%%%%%%%%%
\au_{RR}  \rightarrow 
\frac{1}{2}\left(\frac{R_1}{R_2}\left[\sum_{a}N_a(n^a_1N^a_2+m^a_1M^a_2)n^a_3\right]^2
+ \frac{R_2}{R_1}\left[\sum_{a}N_a(n^a_1M^a_2-m^a_1N^a_2)M^a_3\right]^2\right)\int_0^{\infty} 
dl. \label{FormulaRRtadpoleAn}
\end{equation}

\subsubsection*{RR tadpole cancellation}

The formulas~(\ref{FormulaRRtadpoleKb}), (\ref{FormulaRRtadpoleMS}) and (\ref{FormulaRRtadpoleAn}) provide
two independent RR tadpole cancellation conditions 
corresponding to the fact that the two radii $R_1$ and $R_2$ of $T^2_3$ are not constrained by the
symmetries of the compactification:
\begin{equation}
\begin{aligned} 
& \sum_a\left(n^a_1 N^a_2+m^a_1 M^a_2\right)n^a_3 N_a = 32,\\
& \sum_a\left(n^a_1 M^a_2-m^a_1 N^a_2\right)M^a_3 N_a = \frac{32}{4^b},\\
\end{aligned}\label{EqRRtadpoles}
\end{equation}
where again $b=0,1/2$ corresponds to the two different allowed shapes
of $T^2_3$ and the abbreviations~(\ref{EqWrappingAbb}) have been used.

This matches by comparison with the supersymmetric non-chiral model in~\cite{Forste:2000hx} 
 when taking into account that $\OR$ and $\OR\Theta$ invariant D6-branes occur only in orbits of 
two different cycles (cp. figure~\ref{FigureO6planes})
instead of four in the generic case and thus 
contribute only half 
of the amount of an arbitrary D6-brane to the RR-tadpole cancellation conditions.

\section{General open spectrum} \label{chiralspectrum}

\subsubsection*{Chiral spectrum}

The massless closed string spectrum consists of 57 chiral and one vector multiplet for $b=0$ and of 
47 chiral plus eleven vectormultiplets for $b=1/2$~\cite{Forste:2000hx}. A consistent choice of the 
Chan Paton projection matrices for $a$ and $a'$ simultaneously is provided by
\begin{equation}
\begin{aligned}
%%%%%%%%%%%%%%%%%%%%%%%%%%%%%%%%%%%%%%%%%%%%%%%%%%%%%
\gamma^a_{\Theta} &= \left(\begin{array}{cccc}
e^{i\pi/4}\unity_{N_a/2}  & 0 & 0 & 0\\
0 & e^{-i\pi/4}\unity_{N_a/2} & 0 & 0\\
0 & 0 & e^{-i\pi/4}\unity_{N_a/2} & 0\\
0 & 0 & 0 & e^{i\pi/4}\unity_{N_a/2}
\end{array}\right),\\
%%%%%%%%%%%%%%%%%%%%%%%%%%%%%%%%%%%%%%%%%%%%%%%%%%%%
\gamma^a_{\omega} & = \left(\begin{array}{cccc}
0 &  \unity_{N_a/2} & 0 & 0\\
-\unity_{N_a/2} & 0 & 0 & 0\\
0 & 0 & 0 &  \unity_{N_a/2}\\
0 & 0 & -\unity_{N_a/2} & 0
\end{array}\right),  \qquad
%%%%%%%%%%%%%%%%%%%%%%%%%%%%%%%%%%%%%%%%%%%%%%%%%%%%
\gamma^a_{\OR} = \left(\begin{array}{cccc}
0 & 0 & \unity_{N_a/2} & 0\\
0 & 0 & 0 &  \unity_{N_a/2}\\
\unity_{N_a/2} & 0 & 0 & 0\\
0 &  \unity_{N_a/2} & 0 & 0
\end{array}\right).
%%%%%%%%%%%%%%%%%%%%%%%%%%%%%%%%%%%%%%%%%%%%%%%%%%%%
\end{aligned}   \label{EqChanPatonMatrices}
\end{equation}
The matrices are chosen to be of identical shape for all kinds of D$6_a$-branes. 
The corresponding Chan Paton matrices for the $\Z_4$ rotated branes $(\Theta a)$ are obtained
from the consistency conditions in the M\"obius strip amplitude calculation 
$\gamma_{\OR\Theta^{2k+1}\omega^l}^{T,(\Theta a)'} =\pm \gamma^a_{\OR\Theta^{2k+1}\omega^l}$.
Note that the 
$\OR \times \Z_2 \times \Z_2$ substructure is selected such that it matches the one 
in~\cite{Cvetic:2001tj}. The resulting massless spectra for the cases with
local RR charge cancellation in the compact space, however, coincide with the old 
non-chiral $\OR \times \Z_4 \times \Z_2$ models~\cite{Forste:2000hx} where a different set of matrices 
was used.

The generic gauge group supported by a stack of $N_a$ identical D6-branes is 
broken down by the $\Z_2 \times \Z_2$ symmetry from the initial $U(N_a)$ factor to a subgroup
$U(N_a/2)$. 
If the D6-branes are their own mirror images under ${\cal R}$, the gauge group is further reduced to 
$Sp(N_a/2)$.~\footnote{We follow the convention where the adjoint representation of $Sp(N_a/2)$ 
has $\frac{1}{2}\frac{N_a}{2}(\frac{N_a}{2}+1)$ degrees of freedom.} On a generic representation of
a unitary gauge factor, the antiholomorphic involution ${\cal R}$ acts as complex conjugation.

\begin{figure}
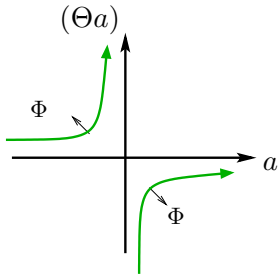

\begin{center}
\input Z4xZ2branerecomb.pstex_t
\end{center}
\caption{Schematic picture of the recombination of a brane $a$ and its $\Z_4$ image $(\Theta a)$ triggered by a
supermultiplet $\Phi$ in the adjoint representation.}
\label{FigureZ4xZ2branerecomb}
\end{figure}

The $aa$ strings provide three chiral multiplets in the adjoint representation of $U(N_a/2)$
which are associated to 
separating identical D6-branes parallely by a distance on the three two-tori $T^2_j$. The $a(\Theta a)$ sector is
supersymmetric as well. It provides a chiral multiplet in the adjoint representation of $U(N_a/2)$
at each intersection point on $T^2_1 \times T^2_2$ which is associated to 
recombining the brane $a$ and its $\Z_4$ image $(\Theta a)$ and
moving the recombined brane away from the intersection point~\cite{Blumenhagen:2002gw}. 
The situation is schematically depicted in figure~\ref{FigureZ4xZ2branerecomb}.
%%%%%%%%%%%%%%%%%%%%%%%%%%%%%%%%%%%%%%%%%%%%%
%%%%%%%%%%%%%%%%%%%%%%%%%%%%%%%%%%%%%%%%%%%%

For arbitrary choices of wrapping numbers and radii $R_1$ and $R_2$,
the remaining sectors $aa'$, $a(\Theta a)'$, $ab \ldots$ are 
non-supersymmetric. The general chiral spectrum is displayed in table~\ref{TableGenSpectrum}.
%%%%%%%%%%%%%%%%%%%%%%%%%%%%%%%%%%%%%%%%%%%%%%%%%
\renewcommand{\arraystretch}{1.3}
\begin{table}[ht]
  \begin{center}
    \begin{equation*}
      \begin{array}{|c|c|c|} \hline
                                % "Uberschrift
        \multicolumn{3}{|c|}{\rule[-3mm]{0mm}{8mm} 
\text{\bf General chiral spectrum}} \\ \hline\hline
       & \text{multiplicity} & \text{representation}
\\ \hline\hline
aa  & & U(N_a/2)\\
 &  3+\frac{1}{2}((n^a_1)^2+(m^a_1)^2)((N^a_2)^2+(M^a_2)^2) & {\bf Adj}_a \\
ab  & I_{ab} + I_{ a(\Theta b)} & ({\bf F}_a,\ov{\bf F}_b)\\
ab' & I_{ab'} + I_{a(\Theta b)'} & ({\bf F}_a,{\bf F}_b)\\
%%%%%%%%%%%
aa' & \frac{1}{2}(I_{aa'}+I_{a(\Theta a)'})  & {\bf Sym}_a\\
&-\frac{1}{2}(I^{\OR\Theta}_{a(\Theta a)'}+I^{\OR\Theta^3}_{a(\Theta a)'}
+I^{\OR\Theta\omega}_{a(\Theta a)'}+I^{\OR\Theta^3\omega}_{a(\Theta a)'})\!\! & \\
%%%%%%%%%%%%%%%%%%%%%%%%%%%%
 &  \frac{1}{2}(I_{aa'}+I_{a(\Theta a)'})  & {\bf Anti}_a\\
&+\frac{1}{2}(I^{\OR\Theta}_{a(\Theta a)'}+I^{\OR\Theta^3}_{a(\Theta a)'} 
+I^{\OR\Theta\omega}_{a(\Theta a)'}+I^{\OR\Theta^3\omega}_{a(\Theta a)'})\!\! & 
\\\hline 
      \end{array}
    \end{equation*}
  \end{center}
\caption{General chiral spectrum for the  $T^6/(\OR \times \Z_4 \times\Z_2)$ orientifold. 
${\bf F}_a = {\bf N_a/2}$ denotes the fundamental representation of 
$U(N_a/2)$. ${\bf Adj}_a$, ${\bf Sym}_a$ and ${\bf Anti}_a$ denote the adjoint, symmetric and 
antisymmetric representations. The explicit
expressions for the intersection numbers are given in appendix~\ref{AppIntersectionNumbers} 
equation~(\ref{EqAppInters}).}
\label{TableGenSpectrum}
\end{table}
%%%%%%%%%%%%%%%%%%%%%%%%%%%%%%%%%%%%%%%%%%%%%%%%%
%%%%%%%%%%%%%%%%%%%%%%%%%%%%%%%%%%%%%%%%%%%%%%%%%
The chiral fermions are accompanied by a bunch of scalar pseudo-superpartners 
with the same gauge quantum numbers 
whose masses 
depend on the six wrapping numbers $n^a_i,m^a_i$, 
the radii $R_1, R_2$ and shape $b=0,1/2$ of the third torus $T^2_3$.

The four kinds of D6-branes which lie on top of the O6-planes correspond to the so called `filler branes' 
of the $T^6/(\Z_2 \times \Z_2)$ model in~\cite{Cvetic:2002pj} plus their $\Z_4$ images. They provide for 
symplectic gauge factors. Their intersections among each 
other provide only for non-chiral matter and are supersymmetric. Their wrapping numbers are 
explicitly listed in appendix~\ref{AppIntersectionNumbers}, equation~(\ref{AppWrapFillerBranes}).
The intersections of `filler branes' with arbitrary D6-branes yield the contribution to the chiral
spectrum displayed in table~\ref{TableSympSpectrum}.
%%%%%%%%%%%%%%%%%%%%%%%%%%%%%%%%%%%%%%%%%%%%%%%%%
\renewcommand{\arraystretch}{1.3}
\begin{table}
  \begin{center}
    \begin{equation*}
      \begin{array}{|c|c|c|} \hline
                                % "Uberschrift
        \multicolumn{3}{|c|}{\rule[-3mm]{0mm}{8mm} 
\text{\bf General chiral spectrum containing `filler branes' $c$}} \\ \hline\hline
       & \text{multiplicity} & \text{representation}
\\ \hline\hline
cc  & & Sp(N_c/2)\\
 &  3+2 & {\bf Anti}_{c} \\
cb  & I_{cb} + I_{ c(\Theta b)}  & ({\bf F}_{c},\ov{\bf F}_b)\\\hline 
      \end{array}
    \end{equation*}
  \end{center}
\caption{General contribution to the chiral spectrum for the  $T^6/(\OR \times \Z_4 \times\Z_2)$ orientifold from
sectors including `filler branes' $c$ and branes $b$ with arbitrary positions. The explicit
expressions for the intersection numbers are given in appendix~\ref{AppIntersectionNumbers}, equation~(\ref{AppIntersFillerBranes}).}
\label{TableSympSpectrum}
\end{table}
%%%%%%%%%%%%%%%%%%%%%%%%%%%%%%%%%%%%%%%%%%%%%%%%%
The vanishing of the chiral anomaly can be checked by imposing the RR tadpole cancellation 
conditions~(\ref{EqRRtadpoles}) bearing in mind that the `filler branes' contribute only  half the
amount to the tadpole equations since they are their own ${\cal R}(\Theta)$ images.

\subsubsection*{Green Schwarz mechanism}

In general, the models provide several $U(1)$ factors. As in the $T^6/\Z_3$ orbifold model 
in~\cite{Blumenhagen:2001te}, the Green Schwarz mechanism is mediated by the two-form ${}^{(4)}B_2$ 
which is 
obtained by dimensional reduction from the ten dimensional two-form ${}^{(10)}C_2$. The dual four dimensional
axion ${}^{(4)}C_0$ is given by ${}^{(4)}C_0=\int_{T^6}{}^{(10)}C_6$. The three other dual pairs 
of RR forms present in 
the toroidal case~\cite{Ibanez:2001nd} are projected out by the orbifold symmetry.
The coupling of a single D$6_a$-brane to the RR forms is the same as in the toroidal case~\cite{Ibanez:2001nd}.
The orbifold symmetry is taken into account by summing over all four D6-branes in the orbit of brane $a$.
%\begin{equation}
%\begin{aligned}
%\int_{\mathbb{R}^{1,3}}\sum_a N_a m^1_a M^2_a M^3_a B_2 \wedge F_a,\\
%\int_{\mathbb{R}^{1,3}}\sum_b  n^1_b N^2_b n^3_b C_0 \wedge F_b\wedge F_b.
%\end{aligned}\label{EqGeneralGScouplings}
%\end{equation}
The resulting couplings are
\begin{equation}
\begin{aligned}
\int_{\mathbb{R}^{1,3}}\sum_a N_a (n^1_a N^2_a+m^1_a M^2_a) M^3_a B_2 \wedge F_a,\\
\int_{\mathbb{R}^{1,3}}\sum_b  (n^1_b N^2_b+m^1_b M^2_b) n^3_b C_0 \wedge F_b\wedge F_b.
\end{aligned}\label{EqOrbitsGScouplings}
\end{equation}
The first equation in~(\ref{EqOrbitsGScouplings}) can give mass to one $U(1)$ factor, even if it is
anomaly-free~\cite{Ibanez:2001nd}.

\subsubsection*{Non-chiral states}

Non-chiral massless states arise from intersections of D6-branes parallel on at least 
one two-torus and from intersections of `filler branes' $c_i$ and $c_j$. These states are listed in 
table~\ref{TableNonchiralSpectrum}.
%%%%%%%%%%%%%%%%%%%%%%%%%%%%%%%%%%%%%%%%%%%%%%%%%
\renewcommand{\arraystretch}{1.3}
\begin{table}
  \begin{center}
    \begin{equation*}
      \begin{array}{|c|c|c|} \hline
                                % "Uberschrift
        \multicolumn{3}{|c|}{\rule[-3mm]{0mm}{8mm} 
\text{\bf Non-chiral spectrum}} \\ \hline\hline
       & \text{multiplicity} & \text{representation}\\ \hline\hline
ab  & \prod_{i\neq i_0}I^i_{ab} + \prod_{i\neq i_0}I^i_{ a(\Theta b)}  
& ({\bf F}_a,\ov{\bf F}_b) + (\ov{\bf F}_a,{\bf F}_b)  \\\hline 
ab'  & \prod_{i\neq i_0}I^i_{ab'} + \prod_{i\neq i_0}I^i_{ a(\Theta b)'}  
& ({\bf F}_a,{\bf F}_b) + (\ov{\bf F}_a,\ov{\bf F}_b)  \\\hline
ac &  \prod_{i\neq i_0}I^i_{ac} + \prod_{i\neq i_0}I^i_{ a(\Theta c)}   
& ({\bf F}_a,{\bf F}_c) + (\ov{\bf F}_a,{\bf F}_c)  \\\hline 
%%%%%%%%%%%%%%%%%%%%
c_1c_2 & 3 \cdot 4^b & ({\bf F}_{c_1},{\bf F}_{c_2})  \\\hline 
c_1c_3 & 2           & ({\bf F}_{c_1},{\bf F}_{c_3})  \\\hline 
c_1c_4 &  4^b        & ({\bf F}_{c_1},{\bf F}_{c_4})  \\\hline 
c_2c_3 &  4^b        & ({\bf F}_{c_2},{\bf F}_{c_3})  \\\hline 
c_2c_4 & 2           & ({\bf F}_{c_2},{\bf F}_{c_4})  \\\hline 
c_3c_4 & 3 \cdot 4^b & ({\bf F}_{c_3},{\bf F}_{c_4})  \\\hline 
      \end{array}
    \end{equation*}
  \end{center}
\caption{Massless non-chiral spectrum for the  $T^6/(\OR \times \Z_4 \times\Z_2)$ orientifold with
sectors including `filler branes'. In addition, representations $({\bf Anti}_a+\ov{\bf Anti}_a)$ and 
$({\bf Sym}_a+\ov{\bf Sym}_a)$ occur at intersection of $a$ with $a'$ and $(\Theta a)'$ if they are parallel
on at least one torus.}
\label{TableNonchiralSpectrum}
\end{table}
%%%%%%%%%%%%%%%%%%%%%%%%%%%%%%%%%%%%%%%%%%%%%%%%% 
The part of the non-chiral spectrum which involves only `filler branes' is supersymmetric and can easily 
be seen to be identical to the one computed in~\cite{Forste:2000hx} for the special case
$N_{c_1}=N_{c_2}=N_{c_3}=N_{c_4}=16/4^b$.

\section{NSNS tadpoles} \label{NSNStadpoles}

The computation of NSNS and NSNS$(-1)^F$ contributions to the tadpoles goes along the same lines as 
the one of the RR tadpoles. One obtains a sum of perfect squares, 
\begin{equation}
\left(\sum_a \frac{L^a_1L^a_2L^a_3}{R_AR_Bv}-16 \sqrt{2}( u+\frac{1}{4^b}\frac{1}{u})\right)^2
+\left(\sum_a \frac{L^a_1L^a_2}{R_AR_Bv}\frac{(n^a_3R_1)^2-(M^a_3R_2)^2}{L^a_3}
-16\sqrt{2}( u-\frac{1}{4^b}\frac{1}{u})\right)^2
\nonumber
\end{equation}
where $v=\sqrt{R_1R_2}$ and $u=\sqrt{R_1/R_2}$ are the 
square roots of the volume and the imaginary part of the complex structure of $T^2_3$.
Using the notation 
\begin{equation}
%\begin{aligned}
{\cal L}^a_3(u) = \sqrt{(n^a_3u)^2+ \left(\frac{M^a_3}{u}\right)^2}, \qquad
vu\partial_u{\cal L}^a_3(u) = \frac{(n^a_3R_1)^2-(M^a_3R_2)^2}{L^a_3},
\nonumber%\end{aligned}
\end{equation}
the NSNS tadpoles can be rewritten as
\begin{equation*}
\left(\sum_a \frac{L^a_1L^a_2{\cal L}^a_3}{R_AR_B}-16\sqrt{2}( u+\frac{1}{4^b}\frac{1}{u})\right)^2
+\left(u\partial_u\left[\sum_a
    \frac{L^a_1L^a_2{\cal L}^a_3}{R_AR_B}
-16\sqrt{2}( u+\frac{1}{4^b}\frac{1}{u})\right]\right)^2.
\end{equation*}
The term in the first bracket is proportional to the overall volume of the D6-branes minus the volume 
of the O6-planes and therefore the dilaton tadpole. The term in the second bracket gives the 
complex structure tadpole for $T^2_3$ since it is proportional to the derivative with respect to $u$.

From this identification of the NSNS tadpoles, the scalar potential is deduced to be
\begin{equation}
V(\Phi_s,u)=e^{-\Phi_s}
\left(\sum_{a=1}^K  N_a \frac{\prod_{i=1}^3 L^i_a}{R_AR_BR_1}-16\sqrt{2}(u+\frac{1}{4^b}\frac{1}{u})\right).
\label{EqScalarPotential}
\end{equation}
Using the relation~(\ref{AppAnglesWrappings}) in appendix~\ref{AppSusycond} between wrapping numbers and 
the tangent of the angles $\varphi^a_i$
with respect to the real axes,
the compact volume of a D$6_a$-brane can in general be rewritten as
\begin{equation*}
\left(\prod_{i=1}^3 L^a_i \right)^2=
\left(R_AR_BR_1n^a_1\frac{N^a_2}{\sqrt{2}}n^a_3\right)^2\left(
\left(1-\sum_{i=1}^3\tan\varphi^a_i\tan\varphi^a_{i+1}\right)^2
+\left(\sum_{i=1}^3\tan\varphi^a_i-\prod_{i=1}^3\tan\varphi^a_i\right)^2\right).
\end{equation*}
Since the necessary condition for preserving supersymmetry, $\sum_{i=1}^3 \varphi^a_i=0$, 
in terms of the tangents reads
\begin{equation}
\sum_{i=1}^3 \tan\varphi^a_i=\prod_{i=1}^3 \tan\varphi^a_i,   \label{EqSUSYtan}
\end{equation}
it is quite obvious that the scalar potential~(\ref{EqScalarPotential}) vanishes iff the
whole D6-brane configuration preserves supersymmetry.

\section{Chiral supersymmetric models} \label{SUSYmodels}

One of the motivations to study type IIA compactifications on $T^6/(\Z_4 \times \Z_2)$ 
with intersecting D6-branes consists in the quest for phenomenologically appealing 
chiral spectra. It turns out to be a difficult task to find general classes of models which satisfy~(\ref{EqSUSYtan}) 
and are free of symmetric representations of a
unitary gauge factor. In general, six wrapping numbers per D6-brane orbit are only constrained by 
these two conditions. The only relevant modulus of the compactification is the ratio of radii $R_1/R_2$
which is fixed by a single supersymmetric D6-brane orbit at non-trivial angle on $T^2_3$ 
together with the discrete 
quantity $b=0$ or $1/2$.

\begin{figure}
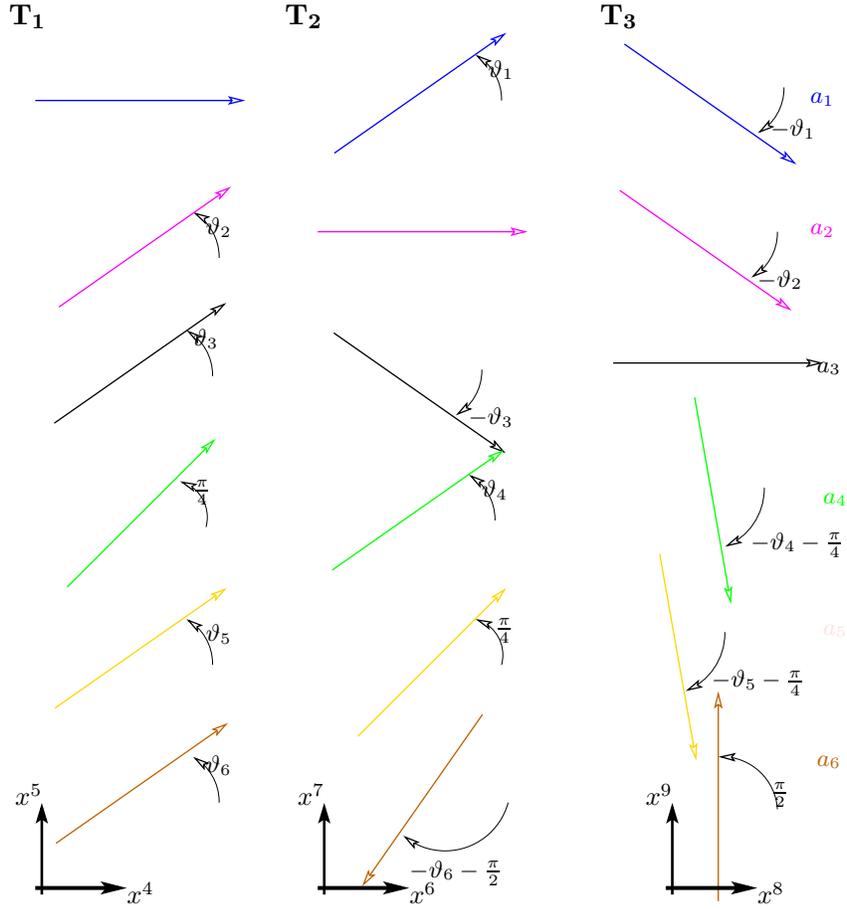

\begin{center}
\input Z4xZ2specialbranes.pstex_t
\end{center}
\caption{Special supersymmetric D6-brane configurations. Only one representant of each orbit 
containing four distinct D6-branes
is depicted for the sake of clarity.
The lattice on $T^2_1 \times T^2_2$ is {\bf AB}, $T^2_3$ can be rectangular or tilted.}
\label{FigureSpecialbranes}
\end{figure}

In order to start the search systematically, we restrict to the cases where
the D6-branes lie on top of an O6-plane on one two-torus. By virtue of supersymmetry, 
these D6-branes are specified by only one non-trivial angle $\vartheta$. There exist six different orbits
of this kind with respect to $\Theta$ and ${\cal R}$. Each orbit is characterized by one of 
the following D6-branes,
\begin{equation}
\begin{aligned}
%%%%%%%%%%%
a_1: \qquad &
\varphi_1= 0,
\varphi_2=\vartheta_1,
\varphi_3=-\vartheta_1,\\
%%%%%%%%%%%%%%%%%%%%%%%%%%%%%%%%
a_2: \qquad &
\varphi_1= \vartheta_2,
\varphi_2= 0,
\varphi_3=-\vartheta_2,\\
%%%%%%%%%%%%%%%%%%%%%%%%%%%%%%%%
a_3: \qquad &
\varphi_1=\vartheta_3, 
\varphi_2=-\vartheta_3,
\varphi_3= 0,\\
%%%%%%%%%%%%%%%%%%%%%%%%%%%%%%%%
a_4: \qquad &
\varphi_1=\frac{\pi}{4},
\varphi_2=\vartheta_4, 
\varphi_3=-\vartheta_4-\frac{\pi}{4},\\
%%%%%%%%%%%%%%%%%%%%%%%%%%%%%%%%
a_5: \qquad &
\varphi_1=\vartheta_5,  
\varphi_2=\frac{\pi}{4},
\varphi_3=-\vartheta_5-\frac{\pi}{4},\\
%%%%%%%%%%%%%%%%%%%%%%%%%%%%%%%%
a_6: \qquad &
\varphi_1= \vartheta_6,  
\varphi_2=-\vartheta_6-\frac{\pi}{2},
\varphi_3=\frac{\pi}{2}.
%%%%%%%%%%%%%%%%%%%%%%%%%%%%%%%%
\end{aligned}\label{EqAngles6SUSYcases}
\end{equation}
The corresponding 3-cycles are depicted in figure~\ref{FigureSpecialbranes}.
The necessary conditions for preserving supersymmetry in each case in terms of wrapping numbers are
explicitly listed in appendix~\ref{AppSusycond}, equation~(\ref{AppWrappings6SUSYcases}).
%%%%%%%%%%%%%%%%%%%%%%%%%%%%%%%%%%%%%%%%%%%%%
%%%%%%%%%%%%%%%%%%%%%%%%%%%%%%%%%%%%%%%%%%%%

In the cases $a_3 \ldots a_6$, the choices of angles $\vartheta_i =0,\frac{\pi}{4}$ ($\text{mod}\frac{\pi}{2}$) 
correspond to selecting one of the supersymmetric `filler branes' $c_j (j=1 \ldots 4)$, and the D6-brane orbit again 
contains only two distinct 3-cycles. 

 In the cases of D6-branes $a_1$ and $a_2$, choosing $\vartheta_i =\frac{\pi}{4}$ leads to a non-trivial 
supersymmetric configuration. The following identities hold independently of the 
values of the radii $R_1, R_2$ and $b$ as long as 
the angles are chosen to be $\vartheta_1=\vartheta_2=\frac{\pi}{4}$:
\begin{align*}
%%%%%%%%%%%
I_{a_1 a_2}+I_{a_1(\Theta a_2)} &= 0,\\
I_{a_1 a_2'}+I_{a_1(\Theta a_2)'} &= 0,\\
I^{a_1}_{Sym} = -I^{a_1}_{Anti} &= 2 n^{a_1}_3 \left( -\frac{1}{4^b}+\frac{R_1}{R_2} \right),\\
I^{a_2}_{Sym} = -I^{a_2}_{Anti} &= 4 n^{a_2}_3 \left( -\frac{1}{4^b}+\frac{R_1}{R_2} \right).\\
%%%%%%%%%%%%%%%%%%%%%%%%%%%%%%%%
\end{align*}
This means that there is no antisymmetric or symmetric representation of a chiral field associated to the stacks 
$a_1$ and $a_2$, as long as 
\begin{equation*}
\frac{R_1}{R_2}=\frac{1}{4^b}
\end{equation*}
is fulfilled, which is the condition for obtaining the rectangular {\bf A} and {\bf B} lattice for 
$b=0,1/2$ respectively.

Implementing the constraint $\vartheta_1=\vartheta_2 =\frac{\pi}{4}$  together with no symmetric representation 
of the gauge group fixes the wrapping numbers of the first two kinds of special D6-branes completely up to
the choice $b=0,1/2$ of the lattice $T^2_3$,
\begin{equation}
\begin{aligned}
%%%%%%%%%%%
\text{brane:}\qquad  & (n_1,m_1) \times (N_2,M_2) \times (n_3,M_3)\\
A_1: \qquad & (1,0)  \times (1,-1) \times (1, -\frac{1}{4^b}),\\
A_2: \qquad & (1,1)  \times (2,0) \times (1, -\frac{1}{4^b}).
%%%%%%%%%%%%%%%%%%%%%%%%%%%%%%%%
\end{aligned}\label{EqbranesA1A2}
\end{equation}
Constraining the D6-brane set-up now to consist of only these two non-trivial kinds of D6-branes $A_1, A_2$ and 
the four kinds of `filler branes'  $c_j$, the tadpole cancellation conditions read
\begin{equation}
32 =N_{A_1} +2 N_{A_2}+4^b\left(N_{c_1} +N_{c_3}\right)=N_{A_1} +2 N_{A_2}+4^b\left(N_{c_2} +N_{c_4}\right).
\nonumber
\end{equation}
The resulting gauge group of this set-up is $U(\frac{N_{A_1}}{2}) \times U(\frac{N_{A_2}}{2}) 
\times Sp(\frac{N_{c_1}}{2}) 
\times Sp(\frac{N_{c_2}}{2}) 
\times Sp(\frac{N_{c_3}}{2})\times Sp(\frac{N_{c_4}}{2})$,
 and the corresponding chiral spectrum 
consists only of bifundamental states,
\begin{equation}
\begin{aligned}
&1 \times \left[ ({\bf F}_{A_1},{\bf F}_{c_1})+(\ov{\bf F}_{A_1},{\bf F}_{c_2})
+({\bf F}_{A_1},{\bf F}_{c_3})+(\ov{\bf F}_{A_1},{\bf F}_{c_4})\right]\\
+ & 2  \times \left[ ({\bf F}_{A_2},{\bf F}_{c_1})+(\ov{\bf F}_{A_2},{\bf F}_{c_2})
+({\bf F}_{A_2},{\bf F}_{c_3})+(\ov{\bf F}_{A_2},{\bf F}_{c_4})\right].
\end{aligned}
\nonumber
\end{equation}
As already mentioned in section~\ref{chiralspectrum}, the notation of symplectic gauge factors
is such that $\frac{N_{c_j}}{2} \in 2 \mathbb{N}$ is required.

\subsubsection*{A supersymmetric Pati-Salam model with four generations}

The general class of models described in the last paragraph can be used to construct explicit semi-realistic
chiral spectra. A supersymmetric Pati-Salam model with four generations is obtained by choosing the 
numbers of identical D6-branes to be 
$N_{A_2}=8$, $N_{c_1}=N_{c_2}=N_{c_3}=N_{c_4}=4$  and the tilted torus, i.e. $b=1/2$. The resulting gauge group is
\begin{equation}
U(4) \times Sp(2)^4 
\label{EqPS4gaugegroup}
\end{equation}
with the chiral matter content
\begin{equation}
2  \times \left[ ({\bf 4},{\bf 2}_1)+(\ov{\bf 4},{\bf 2}_2)
+({\bf 4},{\bf 2}_3)+(\ov{\bf 4},{\bf 2}_4)
\right].
\label{EqPS4chiralspectrum}
\end{equation}
In this case, since $Sp(2) \simeq SU(2)$, the gauge group $U(4) \times SU(2)_L \times SU(2)_R$ 
is obtained by identifying $SU(2)_1 \simeq SU(2)_3$ and $SU(2)_2 \simeq SU(2)_4$ 
via brane recombination. The required kind of Higgs 
effect is commented on below. The unitary factor $U(4)$ is, furthermore, broken down to $U(3) \times U(1)$ by
separating the stack of branes $A_2$ at least on one two-torus, e.g. on $T^2_3$, 
into two parallel sets with $N^{(1)}_{A_2}=6$, $N^{(2)}_{A_2}=2$.
In the T-dual type IIB orientifold, the breaking of the gauge group is triggered by a Wilson line
on the corresponding two-torus. From a 
field theoretic point of view, the three adjoint fields originating from  the  $aa$ sector
parameterize the brane positions on the three two-tori $T^2_j$. A suitable vev. for one of the adjoint 
scalars leads 
to the required breaking of the 
gauge group. This mechanism can always occur, since these adjoint scalars
are flat directions of the compactification. 

Out of the two $U(1)$ factors which emerge from the decomposition 
$U(4)\rightarrow U(3) \times U(1)_2 \rightarrow SU(3) \times U(1)_1 \times U(1)_2$ 
in this model, the linear combination 
\begin{equation}
U(1)_{B-L}=\frac{1}{3}(U(1)_1-3U(1)_2)
\end{equation}
is massless and anomaly-free. The quantum numbers of the chiral fields under $U(1)_{B-L}$ are the
correct ones for a model containing right handed neutrinos. The second $U(1)$ factor is also 
anomaly-free, but acquires a mass through the Green Schwarz mechanism in equation~(\ref{EqOrbitsGScouplings}).

\subsubsection*{A supersymmetric Pati-Salam model with three generations}

Starting again with the set of D6-branes $A_1, A_2, c_1,c_2,c_3,c_4$, it is possible to construct models with 
three generations but two $U(4)$ factors.
More concretely, the choice $N_{A_1}=N_{A_2}=8$, $N_{c_1}=N_{c_2}=4$ with $b=1/2$ leads to a left-right  
symmetric Pati - Salam  model with initial gauge group
\begin{equation}
U(4)_1 \times U(4)_2 \times Sp(2)_1 \times Sp(2)_2
\label{EqPS3gaugegroup}
\end{equation}
and  chiral spectrum
\begin{equation}
1 \times \left[({\bf 4}_1,{\bf 2}_1)+(\ov{\bf 4}_1,{\bf 2}_2)\right]
+  2  \times \left[ ({\bf 4}_2,{\bf 2}_1)+(\ov{\bf 4}_2,{\bf 2}_2)\right].
\label{EqPS3chiralspectrum}
\end{equation}
Brane recombination of $A_1$ and $A_2$ into a single non-factorizable brane and 
the isomorphism $Sp(2)_1\simeq SU(2)_L$, $Sp(2)_2 \simeq SU(2)_R$ lead to the usual Pati - Salam group 
$U(4) \times SU(2)_L \times SU(2)_R$ with three generations of chiral matter 
$3 \times \left[({\bf 4},{\bf 2}_L)+(\ov{\bf 4},{\bf 2}_R)\right]$.

The discussion of Abelian symmetries is identical to the one for the four generation model, i.e. a separation
on $T^2_3$ 
of the stacks $A_1,A_2$ into parallel sets $N^{(1)}_{A_1}=N^{(1)}_{A_2}=6, N^{(2)}_{A_1}=N^{(2)}_{A_2}=2$ 
leads to the desired gauge symmetry breaking $U(4) \rightarrow SU(3) \times U(1)_{B-L}$ after 
brane recombination.

\subsubsection*{Brane recombination}

In the model under consideration with three generations, 
the two kinds of branes $A_1$, $A_2$  which are supposed to recombine into a single
non-factorizable brane preserve a common ${\cal N}=2$ supersymmetry since they are parallel on the third
two-torus $T^2_3$. In general, a brane recombination will take place if the volume of the complex cycle 
wrapped by the non-factorizable brane is smaller than the the sum of the volumina of the factorizable 
cycles. On the other hand, since the original configuration is supersymmetric in this case, the volume of the final 
brane configuration will be the same as the initial one and the configuration is marginally stable.

The D-term potential for an $U(M_a) \times U(M_b)$ gauge symmetry with $M_a=M_b$ and 
\mbox{${\cal N}=2$} hypermultiplets whose complex scalar components are 
denoted by $H^i_1$, $H^i_2$ in the bifundamental 
representations 
$({\bf F}_a,\ov{\bf F}_b)$ and $(\ov{\bf F}_a,{\bf F}_b)$, respectively, 
vanishes provided that the following equations are fulfilled,
\begin{equation}
\begin{aligned}
\sum_i \left(H^i_1 T^a_G (H^i_1)^{\dagger}-(H^i_2)^{\dagger} T^a_G H^i_2\right) & =0,\\
\sum_i \left((H^i_1)^{\dagger} T^b_G H^i_1-H^i_2 T^b_G (H^i_2)^{\dagger}\right) & =0.
\end{aligned}\label{EqDtermConstraints}
\end{equation}
$i=1 \ldots I_{ab}+I_{a(\Theta b)}$ labels the number of hypermultiplets which is in the intersecting D6-brane
models given by the number of brane intersections. $T^a_G$ and $T^b_G$ label the generators of the gauge factors
$U(M_a)$ and $U(M_b)$, respectively. The gauge indices are summed in such a way that e.g. in the first row
in~(\ref{EqDtermConstraints}), the index of $H^i_1$ pertaining to the first gauge factor $U(M_a)$
is contracted with the generator $T^a_G$ etc. 
Imposing vev.s of the scalars in the hypermultiplets to be proportional to the identity matrix, i.e.
$\langle H^i_j \rangle =h^i_j \unity $ ($j=1,2$), reduces the D-term potential to its
$U(1)$ part~\cite{Blumenhagen:2002gw}
\begin{equation}
V_D=\frac{1}{2g^2}\left( \sum_i h^i_1 \ov{h}^i_1 - \sum_i h^i_2 \ov{h}^i_2 \right)^2.
\label{EqDtermPotential}
\end{equation}
The superpotential contributes to the F-terms via
\begin{equation}
W=\sum_i \left(H^i_1 \Phi_a H^i_2- H^i_1 \Phi_b H^i_2 \right)
\label{EqFtermSuperpotential}
\end{equation}
where $\Phi_a$, $\Phi_b$ are fields in the adjoint representation of the respective gauge factors
and contraction of all gauge indices is understood. Imposing diagonal vev.s as above reduces the requirement of 
vanishing F-term contributions to the potential to 
\begin{equation}
\sum_i h^i_1 h^i_2 =0.
\label{EqFtermConstraint}
\end{equation}
Therefore, a flat direction admitting brane recombination occurs in the presence of at least
two hypermultiplets, e.g. for $h^1_1=h^2_2=h\neq 0$, $h^2_1=h^1_2=0$. This is exactly the case for 
the three generation model with initial gauge group $U(4)_1 \times U(4)_2$ 
where one hypermultiplet is provided by the intersection of $A_1$ with $A_2$ and 
the other one by the intersection of $A_1$ with $(\Theta A_2)$. A generic value for $h$ provides the required 
recombination of $A_1$ and $A_2$ including their respective $\Z_4$ images 
into a single non-factorizable brane with gauge group $U(4)$.

In the model with four generations, brane recombination is supposed to occur among the
$Sp(2)$ factors. The discussion of D-terms in this case is similar to the previous one 
regarding the $Sp(2)$ factors as $SU(2)$s. 
However, due to the absence 
of an Abelian factor, the D-terms vanish identically for vev.s proportional to the identity.
From table~\ref{TableNonchiralSpectrum}, one can read off that there exist always two chiral multiplets 
in the required bifundamental representation of the symplectic gauge factors 
at the intersections $c_1 c_3+c_1 (\Theta c_3)$ and $c_2 c_4+c_2 (\Theta c_4)$. 
These bifundamentals are expected to trigger the 
brane recombination processes $c_1+c_3 \rightarrow c_L$, $c_2+c_4 \rightarrow c_R$
which lead to the final gauge group \mbox{$U(4) \times SU(2)_L \times SU(2)_R$}. A more precise description
of this process would, however, require a more detailed analysis of the F-terms in this ${\cal N}=1$ supersymmetric
sector.

As mentioned in~\cite{Cremades:2003qj}, the $SU(2)_R$ symmetry in both models 
might also be broken down to a $U(1)$ factor
by parallely displacing the brane $c_2$ in the three generation model and
$c_2+c_4=c_R$ in the four generation model from the O6-planes, e.g. on the third two-torus $T^2_3$, 
in such a way that the whole configuration again respects the $\OR$ symmetry.

\section{Conclusions}\label{Conclusions}

In this paper, a type IIA orientifold in the orbifold background $T^6/(\Z_4 \times \Z_2)$ with 
intersecting D6-branes has been studied. The RR tadpole cancellation conditions and the scalar potential at 
disc level have been computed. The general chiral spectrum
has been derived,
the non-chiral part of the low energy spectrum has been considered 
 and two examples of supersymmetric models with a Pati-Salam gauge group  have been given,
 thus proving that it is worthwhile to consider this orientifold model 
in greater detail. In particular, the first supersymmetric and stable
three generation model with right handed fermions but no 
further exotic chiral matter has been found whereas all previous
chiral supersymmetric models from intersecting D6-branes on $T^6/(\Z_2
\times \Z_2)$ and $T^6/\Z_4$ contained some exotic matter. As in the
latter models, the complex structure moduli on two of the three
two-tori are frozen by the $\Z_4 \times \Z_2$ symmetry whereas no such
constraint arises in the $T^6/(\Z_2 \times\Z_2)$ case.

However, many open questions remain. On the one hand, it is worthwhile to search more systematically 
 for (supersymmetric)
models with exactly the chiral MSSM content as well as GUT generalizations. 
Furthermore, 
the non-chiral sectors  need to be studied in full detail in order to understand the exact
mechanisms of gauge- and supersymmetry breaking and brane recombination as well as the low energy 
features of these models. 
Phenomenological aspects of these models including the computation of gauge couplings 
similar to those in~\cite{Cvetic:2002qa} for the $T^6/(\Z_2 \times \Z_2)$ orbifold 
and Yukawa couplings as computed in~\cite{Cremades:2003qj} for the toroidal case also deserve
further study.
Furthermore, the gauge threshold corrections as recently considered
 for intersecting D6-branes 
in~\cite{Lust:2003ky} deserve to be worked out in detail for the model presented in this article.
However, it has to be taken into account that in these two specific
 models, recombined branes wrap non-factorizable cycles and thus a
 generalization of all phenomenological quantities explicitly derived
 for factorizable branes is required.

Last but not least, 
in order to understand M-theory better, 
it is an interesting issue to explore in more detail the type II and heterotic dual models.

\vskip 1cm

\noindent {\bf Acknowledgments}

\noindent 
It am grateful to D.~Cremades, L.~Ib\'{a}\~{n}ez, F.~Marchesano and  A.~Uranga
for helpful discussions and comments. Furthermore, I wish to thank K.~Wendland for 
the computation of the Hodge numbers in this orbifold model.\\
\noindent
This work is supported by the RTN program under contract number HPRN-CT-2000-00148.

%\newpage
%%%%%%%%%%%%%%%%%%%%%%%%%%%%%%%%%%%%%%%%%%%%%%%%%%%%%%%%%%%%%%%%%%%%%
\begin{appendix}
\section{Intersection numbers}\label{AppIntersectionNumbers}
The explicit expressions for the intersection numbers of the orbits of brane $a$ with brane $b$ 
in term of the wrapping numbers are
%%%%%%%%%%%%%%%%%%%%%%%%%%%%%%%%%%%%%%%%
\begin{equation}
\begin{aligned}
I_{ab} + I_{ a(\Theta b)}  &=
\frac{1}{2} (n^a_1 m^b_1-m^a_1 n^b_1) (N^a_2 M^b_2-M^a_2 N^b_2) (-n^a_3 M^b_3+M^a_3 n^b_3) \\
& +\frac{1}{2}(n^a_1 n^b_1+m^a_1 m^b_1) (N^a_2 N^b_2+M^a_2 M^b_2) (-n^a_3 M^b_3+M^a_3 n^b_3),\\
%%%%%%%%%%%%%%
I_{ab'} + I_{a(\Theta b)'} &=
\frac{1}{2} (n^a_1 m^b_1+m^a_1 n^b_1) (N^a_2 M^b_2+M^a_2 N^b_2) (n^a_3 M^b_3+M^a_3 n^b_3) \\
& -\frac{1}{2} (n^a_1 n^b_1-m^a_1 m^b_1) (-N^a_2 N^b_2+M^a_2 M^b_2) (n^a_3 M^b_3+M^a_3 n^b_3),\\
%%%%%%%%%%%%%%%%%%%%%%%%%%%%%%%%  Sym  %%%%
I^a_{Sym}=\frac{1}{2}(I_{aa'}+I_{a(\Theta a)'}) & \\ 
-\frac{1}{2}(I^{\OR\Theta}_{a(\Theta a)'}+I^{\OR\Theta^3}_{a(\Theta a)'} 
+I^{\OR\Theta\omega}_{a(\Theta a)'} +I^{\OR\Theta^3\omega}_{a(\Theta a)'}) &=
%%%%%%%
 -\frac{2}{4^b} (m^a_1 N^a_2-n^a_1 M^a_2) n^a_3
-2(m^a_1 M^a_2+n^a_1 N^a_2) M^a_3 \\
 & +2m^a_1 n^a_1 M^a_2 N^a_2 n^a_3 M^a_3 \\
& + \frac{1}{2}((n^a_1)^2-(m^a_1)^2)((N^a_2)^2-(M^a_2)^2) n^a_3 M^a_3,\\
%%%%%%%%%%%%%%%%%%%%%%%%%%%%%%%%%%% Anti %%%
I^a_{Anti}= \frac{1}{2}(I_{aa'}+I_{a(\Theta a)'}) & \\ 
+\frac{1}{2}(I^{\OR\Theta}_{a(\Theta a)'}+I^{\OR\Theta^3}_{a(\Theta a)'}
+I^{\OR\Theta\omega}_{a(\Theta a)'} +I^{\OR\Theta^3\omega}_{a(\Theta a)'}) &=
%%%%%%%
 \frac{2}{4^b} (m^a_1 N^a_2-n^a_1 M^a_2) n^a_3
+2(m^a_1 M^a_2+n^a_1 N^a_2) M^a_3 \\
 & +2m^a_1 n^a_1 M^a_2 N^a_2 n^a_3 M^a_3 \\
& + \frac{1}{2}((n^a_1)^2-(m^a_1)^2)((N^a_2)^2-(M^a_2)^2) n^a_3 M^a_3,
%%%%%%%%%%%%%%%%%%%%%%%%%%%%%%%%%%
\end{aligned}\label{EqAppInters}
\end{equation}
where the abbreviations~(\ref{EqWrappingAbb}) have been used.

The wrapping numbers of the ${\cal R}\Theta^k$ invariant `filler branes' are 
\begin{equation}
\begin{aligned}
\text{brane:} \qquad & (n_1,m_1) \times (N_2,M_2) \times (n_3,M_3)\\
c_1:  \qquad & (1,0)\times (2,0) \times (4^b,0),\\
c_2:  \qquad & (1,0)\times (0,-2) \times (0,-1),\\
c_3:  \qquad & (1,1)\times (1,1) \times (4^b,0),\\
c_4:  \qquad & (1,1)\times (1,-1) \times  (0,-1).
\end{aligned}\label{AppWrapFillerBranes}
\end{equation}
For intersections of `filler branes' $c$ with arbitrary branes $b$, 
the intersection numbers are given by 
\begin{equation}
I_{cb} + I_{ c (\Theta b)}  
=\left\{
\begin{array}{ll}
-4^b(n^b_1N^b_2+m^b_1M^b_2)M^b_3 & \qquad M^c_3=0,\\
-(m^b_1N^b_2-n^b_1M^b_2)n^b_3 & \qquad M^c_3=-1.
\end{array}
\right.\label{AppIntersFillerBranes}
\end{equation}
%%%%%%%%%%%%%%%%%%%%%%%%%%%%%%%%%%%%%%%%%%%%%%%%%%%%%%%%%%%%%%%%%%%%%%%%%%%%%%%%
%%%%%%%%%%%%%%%%%%%%%%%%%%%%%%%%%%%%%%%%%%%%%%%%%%%%%%%%%%%%%%%%%%%%%%%%%%%%%%%%
\section{Supersymmetry conditions on intersection numbers}\label{AppSusycond}

The angles on the three two-tori with respect to the real axes in terms of the wrapping numbers, 
radii and shape of $T^2_3$ are given by
\begin{equation}
\begin{aligned}
\tan\varphi^a_1 &=\frac{m^a_1}{n^a_1},\\
\tan\varphi^a_2 &=\frac{m^a_2-n^a_2}{m^a_2+n^a_2}=-\frac{M^a_2}{N^a_2},\\
\tan\varphi^a_3 &=\frac{(m^a_3+bn^a_3)R_2}{n^a_3R_1}=\frac{M^a_3R_2}{n^a_3R_1}.
\end{aligned}\label{AppAnglesWrappings}
\end{equation}
Using the relations~(\ref{AppAnglesWrappings}), the supersymmetry conditions~(\ref{EqAngles6SUSYcases}) 
for the specific D6-branes characterized by one non-trivial angle $\vartheta$ only
can be rephrased in terms of wrapping numbers as follows:
\begin{equation}
\begin{aligned}
%%%%%%%%%%%
a_1: \qquad & (n_1,m_1)=(1,0), \qquad M_2=\frac{M_3R_2}{n_3R_1}N_2,\\
%%%%%%%%%%%%%%%%%%%%%%%%%%%%%%%%
a_2: \qquad & (N_2,M_2)=(2,0), \qquad  m_1=-\frac{M_3R_2}{n_3R_1}n_1,\\
%%%%%%%%%%%%%%%%%%%%%%%%%%%%%%%%
a_3: \qquad & (n_3,M_3)=(4^b,0), \qquad m_1=\frac{M_2}{N_2}n_1,\\
%%%%%%%%%%%%%%%%%%%%%%%%%%%%%%%%
a_4: \qquad & (n_1,m_1)=(1,1), \qquad M_2=\frac{n_3R_1+M_3R_2}{n_3R_1-M_3R_2}N_2,\\
%%%%%%%%%%%%%%%%%%%%%%%%%%%%%%%%
a_5: \qquad &(N_2,M_2)=(1,-1), \qquad  m_1=-\frac{n_3R_1+M_3R_2}{n_3R_1-M_3R_2}n_1,\\
%%%%%%%%%%%%%%%%%%%%%%%%%%%%%%%%
a_6: \qquad & (n_3,M_3)=(0,-1), \qquad m_1=-\frac{N_2}{M_2}n_1.
\end{aligned}\label{AppWrappings6SUSYcases}
\end{equation}
The conditions in terms of the wrapping numbers are only necessary conditions. The 
correct orientation of the D6-branes has to be checked separately.

\end{appendix}

%%%%%%%%%%%%%%%%%%%%%%%%%%%%%%%%%%%%%%%%%%%%%%%%%%%%%%%%%%%%%%%%%%%%%%%

\end{document}

%% file: Z4xZ2fixedpoints.pstex_t
\begin{picture}(0,0)%
\includegraphics{Z4xZ2fixedpoints.pstex}%
\end{picture}%
\setlength{\unitlength}{2368sp}%
\begingroup\makeatletter\ifx\SetFigFont\undefined%
\gdef\SetFigFont#1#2#3#4#5{%
  \reset@font\fontsize{#1}{#2pt}%
  \fontfamily{#3}\fontseries{#4}\fontshape{#5}%
  \selectfont}%
\fi\endgroup%
\begin{picture}(12022,8797)(451,-7936)
\put(9751,-3586){\makebox(0,0)[lb]{\smash{\SetFigFont{12}{14.4}{\rmdefault}{\mddefault}{\updefault}\special{ps: gsave 0 0 0 setrgbcolor}${\bf T}_3:{\bf b=0,1/2}$\special{ps: grestore}}}}
\put(3226,-4936){\makebox(0,0)[lb]{\smash{\SetFigFont{12}{14.4}{\rmdefault}{\mddefault}{\updefault}\special{ps: gsave 0 0 0 setrgbcolor}$x^4$\special{ps: grestore}}}}
\put(4501,-1561){\makebox(0,0)[lb]{\smash{\SetFigFont{12}{14.4}{\rmdefault}{\mddefault}{\updefault}\special{ps: gsave 0 0 0 setrgbcolor}$x^7$\special{ps: grestore}}}}
\put(451,-1636){\makebox(0,0)[lb]{\smash{\SetFigFont{12}{14.4}{\rmdefault}{\mddefault}{\updefault}\special{ps: gsave 0 0 0 setrgbcolor}$x^5$\special{ps: grestore}}}}
\put(8101,-4861){\makebox(0,0)[lb]{\smash{\SetFigFont{12}{14.4}{\rmdefault}{\mddefault}{\updefault}\special{ps: gsave 0 0 0 setrgbcolor}$x^6$\special{ps: grestore}}}}
\put(12001,-7861){\makebox(0,0)[lb]{\smash{\SetFigFont{12}{14.4}{\rmdefault}{\mddefault}{\updefault}\special{ps: gsave 0 0 0 setrgbcolor}$x^8$\special{ps: grestore}}}}
\put(9376,-4786){\makebox(0,0)[lb]{\smash{\SetFigFont{12}{14.4}{\rmdefault}{\mddefault}{\updefault}\special{ps: gsave 0 0 0 setrgbcolor}$x^9$\special{ps: grestore}}}}
\put(12001,-2461){\makebox(0,0)[lb]{\smash{\SetFigFont{12}{14.4}{\rmdefault}{\mddefault}{\updefault}\special{ps: gsave 0 0 0 setrgbcolor}$x^8$\special{ps: grestore}}}}
\put(9376,464){\makebox(0,0)[lb]{\smash{\SetFigFont{12}{14.4}{\rmdefault}{\mddefault}{\updefault}\special{ps: gsave 0 0 0 setrgbcolor}$x^9$\special{ps: grestore}}}}
\put(526,-3211){\makebox(0,0)[lb]{\smash{\SetFigFont{12}{14.4}{\rmdefault}{\mddefault}{\updefault}\special{ps: gsave 0 0 1 setrgbcolor}$R_A$\special{ps: grestore}}}}
\put(1801,-4861){\makebox(0,0)[lb]{\smash{\SetFigFont{12}{14.4}{\rmdefault}{\mddefault}{\updefault}\special{ps: gsave 0 0 1 setrgbcolor}$R_A$\special{ps: grestore}}}}
\put(5551,-3511){\makebox(0,0)[lb]{\smash{\SetFigFont{12}{14.4}{\rmdefault}{\mddefault}{\updefault}\special{ps: gsave 0 0 1 setrgbcolor}$R_B$\special{ps: grestore}}}}
\put(5476,-5761){\makebox(0,0)[lb]{\smash{\SetFigFont{12}{14.4}{\rmdefault}{\mddefault}{\updefault}\special{ps: gsave 0 0 1 setrgbcolor}$R_B$\special{ps: grestore}}}}
\put(10801,-2461){\makebox(0,0)[lb]{\smash{\SetFigFont{12}{14.4}{\rmdefault}{\mddefault}{\updefault}\special{ps: gsave 0 0 1 setrgbcolor}$R_1$\special{ps: grestore}}}}
\put(8551,-961){\makebox(0,0)[lb]{\smash{\SetFigFont{12}{14.4}{\rmdefault}{\mddefault}{\updefault}\special{ps: gsave 0 0 1 setrgbcolor}$R_2$\special{ps: grestore}}}}
\put(8476,-6361){\makebox(0,0)[lb]{\smash{\SetFigFont{12}{14.4}{\rmdefault}{\mddefault}{\updefault}\special{ps: gsave 0 0 1 setrgbcolor}$R_2$\special{ps: grestore}}}}
\put(10126,-7936){\makebox(0,0)[lb]{\smash{\SetFigFont{12}{14.4}{\rmdefault}{\mddefault}{\updefault}\special{ps: gsave 0 0 1 setrgbcolor}$R_1$\special{ps: grestore}}}}
\put(1276,-2686){\makebox(0,0)[lb]{\smash{\SetFigFont{12}{14.4}{\rmdefault}{\mddefault}{\updefault}\special{ps: gsave 0 0 1 setrgbcolor}$e^{\scriptstyle 1}_2$\special{ps: grestore}}}}
\put(3076,-4411){\makebox(0,0)[lb]{\smash{\SetFigFont{12}{14.4}{\rmdefault}{\mddefault}{\updefault}\special{ps: gsave 0 0 1 setrgbcolor}$e^{\scriptstyle 1}_1$\special{ps: grestore}}}}
\put(6451,-2986){\makebox(0,0)[lb]{\smash{\SetFigFont{12}{14.4}{\rmdefault}{\mddefault}{\updefault}\special{ps: gsave 0 0 1 setrgbcolor}$e^{\scriptstyle 2}_2$\special{ps: grestore}}}}
\put(6226,-6286){\makebox(0,0)[lb]{\smash{\SetFigFont{12}{14.4}{\rmdefault}{\mddefault}{\updefault}\special{ps: gsave 0 0 1 setrgbcolor}$e^{\scriptstyle 2}_1$\special{ps: grestore}}}}
\put(8626,-361){\makebox(0,0)[lb]{\smash{\SetFigFont{12}{14.4}{\rmdefault}{\mddefault}{\updefault}\special{ps: gsave 0 0 1 setrgbcolor}$e^{\scriptstyle 3}_2$\special{ps: grestore}}}}
\put(11476,-2011){\makebox(0,0)[lb]{\smash{\SetFigFont{12}{14.4}{\rmdefault}{\mddefault}{\updefault}\special{ps: gsave 0 0 1 setrgbcolor}$e^{\scriptstyle 3}_1$\special{ps: grestore}}}}
\put(8551,-5761){\makebox(0,0)[lb]{\smash{\SetFigFont{12}{14.4}{\rmdefault}{\mddefault}{\updefault}\special{ps: gsave 0 0 1 setrgbcolor}$e^{\scriptstyle 3}_2$\special{ps: grestore}}}}
\put(11401,-6961){\makebox(0,0)[lb]{\smash{\SetFigFont{12}{14.4}{\rmdefault}{\mddefault}{\updefault}\special{ps: gsave 0 0 1 setrgbcolor}$e^{\scriptstyle 3}_1$\special{ps: grestore}}}}
\put(1426,-586){\makebox(0,0)[lb]{\smash{\SetFigFont{12}{14.4}{\rmdefault}{\mddefault}{\updefault}\special{ps: gsave 0 0 0 setrgbcolor}${\bf T}_1:{\bf A}$\special{ps: grestore}}}}
\put(5551,-586){\makebox(0,0)[lb]{\smash{\SetFigFont{12}{14.4}{\rmdefault}{\mddefault}{\updefault}\special{ps: gsave 0 0 0 setrgbcolor}${\bf T}_2:{\bf B}$\special{ps: grestore}}}}
\end{picture}

%% file: Z4xZ2Oplanes.pstex_t
\begin{picture}(0,0)%
\includegraphics{Z4xZ2Oplanes.pstex}%
\end{picture}%
\setlength{\unitlength}{2368sp}%
\begingroup\makeatletter\ifx\SetFigFont\undefined%
\gdef\SetFigFont#1#2#3#4#5{%
  \reset@font\fontsize{#1}{#2pt}%
  \fontfamily{#3}\fontseries{#4}\fontshape{#5}%
  \selectfont}%
\fi\endgroup%
\begin{picture}(9044,2638)(279,-5183)
\put(8326,-4936){\makebox(0,0)[lb]{\smash{\SetFigFont{8}{9.6}{\rmdefault}{\mddefault}{\updefault}\special{ps: gsave 0 0 0 setrgbcolor}{\bf $\omega$}\special{ps: grestore}}}}
\put(301,-2761){\makebox(0,0)[lb]{\smash{\SetFigFont{12}{14.4}{\rmdefault}{\mddefault}{\updefault}\special{ps: gsave 0 0 0 setrgbcolor}$\mbox{\bf T}_{\mbox{\scriptsize \bf 1}}$\special{ps: grestore}}}}
\put(3601,-2761){\makebox(0,0)[lb]{\smash{\SetFigFont{12}{14.4}{\rmdefault}{\mddefault}{\updefault}\special{ps: gsave 0 0 0 setrgbcolor}$\mbox{\bf T}_{\mbox{\scriptsize \bf 2}}$\special{ps: grestore}}}}
\put(6901,-2761){\makebox(0,0)[lb]{\smash{\SetFigFont{12}{14.4}{\rmdefault}{\mddefault}{\updefault}\special{ps: gsave 0 0 0 setrgbcolor}$\mbox{\bf T}_{\mbox{\scriptsize \bf 3}}$\special{ps: grestore}}}}
\put(1876,-3361){\makebox(0,0)[lb]{\smash{\SetFigFont{8}{9.6}{\rmdefault}{\mddefault}{\updefault}\special{ps: gsave 0 0 0 setrgbcolor}{\bf $\Theta$}\special{ps: grestore}}}}
\put(5251,-4561){\makebox(0,0)[lb]{\smash{\SetFigFont{8}{9.6}{\rmdefault}{\mddefault}{\updefault}\special{ps: gsave 0 0 0 setrgbcolor}{\bf $\Theta$}\special{ps: grestore}}}}
\put(5176,-3286){\makebox(0,0)[lb]{\smash{\SetFigFont{8}{9.6}{\rmdefault}{\mddefault}{\updefault}\special{ps: gsave 0 0 0 setrgbcolor}{\bf $\omega$}\special{ps: grestore}}}}
\end{picture}

%% file: Z4xZ2braneimages.pstex_t
\begin{picture}(0,0)%
\includegraphics{Z4xZ2braneimages.pstex}%
\end{picture}%
\setlength{\unitlength}{2171sp}%
\begingroup\makeatletter\ifx\SetFigFont\undefined%
\gdef\SetFigFont#1#2#3#4#5{%
  \reset@font\fontsize{#1}{#2pt}%
  \fontfamily{#3}\fontseries{#4}\fontshape{#5}%
  \selectfont}%
\fi\endgroup%
\begin{picture}(12218,6607)(105,-6923)
\put(9526,-586){\makebox(0,0)[lb]{\smash{\SetFigFont{11}{13.2}{\rmdefault}{\mddefault}{\updefault}\special{ps: gsave 0 0 0 setrgbcolor}${\bf T}_3:{\bf b=1/2}$\special{ps: grestore}}}}
\put(3226,-4936){\makebox(0,0)[lb]{\smash{\SetFigFont{11}{13.2}{\rmdefault}{\mddefault}{\updefault}\special{ps: gsave 0 0 0 setrgbcolor}$x^4$\special{ps: grestore}}}}
\put(4501,-1561){\makebox(0,0)[lb]{\smash{\SetFigFont{11}{13.2}{\rmdefault}{\mddefault}{\updefault}\special{ps: gsave 0 0 0 setrgbcolor}$x^7$\special{ps: grestore}}}}
\put(451,-1636){\makebox(0,0)[lb]{\smash{\SetFigFont{11}{13.2}{\rmdefault}{\mddefault}{\updefault}\special{ps: gsave 0 0 0 setrgbcolor}$x^5$\special{ps: grestore}}}}
\put(8101,-4861){\makebox(0,0)[lb]{\smash{\SetFigFont{11}{13.2}{\rmdefault}{\mddefault}{\updefault}\special{ps: gsave 0 0 0 setrgbcolor}$x^6$\special{ps: grestore}}}}
\put(3076,-4411){\makebox(0,0)[lb]{\smash{\SetFigFont{11}{13.2}{\rmdefault}{\mddefault}{\updefault}\special{ps: gsave 0 0 1 setrgbcolor}$e_1$\special{ps: grestore}}}}
\put(6226,-6286){\makebox(0,0)[lb]{\smash{\SetFigFont{11}{13.2}{\rmdefault}{\mddefault}{\updefault}\special{ps: gsave 0 0 1 setrgbcolor}$e_1$\special{ps: grestore}}}}
\put(1276,-2686){\makebox(0,0)[lb]{\smash{\SetFigFont{11}{13.2}{\rmdefault}{\mddefault}{\updefault}\special{ps: gsave 0 0 1 setrgbcolor}$e_2$\special{ps: grestore}}}}
\put(6451,-2986){\makebox(0,0)[lb]{\smash{\SetFigFont{11}{13.2}{\rmdefault}{\mddefault}{\updefault}\special{ps: gsave 0 0 1 setrgbcolor}$e_2$\special{ps: grestore}}}}
\put(8476,-1561){\makebox(0,0)[lb]{\smash{\SetFigFont{11}{13.2}{\rmdefault}{\mddefault}{\updefault}\special{ps: gsave 0 0 0 setrgbcolor}$x^9$\special{ps: grestore}}}}
\put(8776,-2686){\makebox(0,0)[lb]{\smash{\SetFigFont{11}{13.2}{\rmdefault}{\mddefault}{\updefault}\special{ps: gsave 0 0 1 setrgbcolor}$e_2$\special{ps: grestore}}}}
\put(11626,-3886){\makebox(0,0)[lb]{\smash{\SetFigFont{11}{13.2}{\rmdefault}{\mddefault}{\updefault}\special{ps: gsave 0 0 1 setrgbcolor}$e_1$\special{ps: grestore}}}}
\put(12001,-4936){\makebox(0,0)[lb]{\smash{\SetFigFont{11}{13.2}{\rmdefault}{\mddefault}{\updefault}\special{ps: gsave 0 0 0 setrgbcolor}$x^8$\special{ps: grestore}}}}
\put(1201,-3661){\makebox(0,0)[lb]{\smash{\SetFigFont{9}{10.8}{\rmdefault}{\mddefault}{\updefault}\special{ps: gsave 0 0 0 setrgbcolor}$\Theta$\special{ps: grestore}}}}
\put(6001,-5311){\makebox(0,0)[lb]{\smash{\SetFigFont{9}{10.8}{\rmdefault}{\mddefault}{\updefault}\special{ps: gsave 0 0 0 setrgbcolor}$\Theta$\special{ps: grestore}}}}
\put(1426,-586){\makebox(0,0)[lb]{\smash{\SetFigFont{11}{13.2}{\rmdefault}{\mddefault}{\updefault}\special{ps: gsave 0 0 0 setrgbcolor}${\bf T}_1:{\bf A}$\special{ps: grestore}}}}
\put(5551,-586){\makebox(0,0)[lb]{\smash{\SetFigFont{11}{13.2}{\rmdefault}{\mddefault}{\updefault}\special{ps: gsave 0 0 0 setrgbcolor}${\bf T}_2:{\bf B}$\special{ps: grestore}}}}
\end{picture}

%% file: Z4xZ2branerecomb.pstex_t
\begin{picture}(0,0)%
\includegraphics{Z4xZ2branerecomb.pstex}%
\end{picture}%
\setlength{\unitlength}{1973sp}%
\begingroup\makeatletter\ifx\SetFigFont\undefined%
\gdef\SetFigFont#1#2#3#4#5{%
  \reset@font\fontsize{#1}{#2pt}%
  \fontfamily{#3}\fontseries{#4}\fontshape{#5}%
  \selectfont}%
\fi\endgroup%
\begin{picture}(3258,3503)(7168,-4254)
\put(7501,-2236){\makebox(0,0)[lb]{\smash{\SetFigFont{9}{10.8}{\rmdefault}{\mddefault}{\updefault}\special{ps: gsave 0 0 0 setrgbcolor}$\Phi$\special{ps: grestore}}}}
\put(10426,-2911){\makebox(0,0)[lb]{\smash{\SetFigFont{11}{13.2}{\rmdefault}{\mddefault}{\updefault}\special{ps: gsave 0 0 0 setrgbcolor}$a$\special{ps: grestore}}}}
\put(7876,-1111){\makebox(0,0)[lb]{\smash{\SetFigFont{11}{13.2}{\rmdefault}{\mddefault}{\updefault}\special{ps: gsave 0 0 0 setrgbcolor}$(\Theta a)$\special{ps: grestore}}}}
\put(9226,-3586){\makebox(0,0)[lb]{\smash{\SetFigFont{9}{10.8}{\rmdefault}{\mddefault}{\updefault}\special{ps: gsave 0 0 0 setrgbcolor}$\Phi$\special{ps: grestore}}}}
\end{picture}

%% file: Z4xZ2specialbranes.pstex_t
\begin{picture}(0,0)%
\includegraphics{Z4xZ2specialbranes.pstex}%
\end{picture}%
\setlength{\unitlength}{2171sp}%
\begingroup\makeatletter\ifx\SetFigFont\undefined%
\gdef\SetFigFont#1#2#3#4#5{%
  \reset@font\fontsize{#1}{#2pt}%
  \fontfamily{#3}\fontseries{#4}\fontshape{#5}%
  \selectfont}%
\fi\endgroup%
\begin{picture}(9322,10288)(1,-10133)
\put(7276,-8986){\makebox(0,0)[lb]{\smash{\SetFigFont{10}{12.0}{\rmdefault}{\mddefault}{\updefault}\special{ps: gsave 0 0 0 setrgbcolor}$x^9$\special{ps: grestore}}}}
\put(  1,-61){\makebox(0,0)[lb]{\smash{\SetFigFont{11}{13.2}{\rmdefault}{\mddefault}{\updefault}\special{ps: gsave 0 0 0 setrgbcolor}$\mbox{\bf T}_{\mbox{\scriptsize \bf 1}}$\special{ps: grestore}}}}
\put(3151,-61){\makebox(0,0)[lb]{\smash{\SetFigFont{11}{13.2}{\rmdefault}{\mddefault}{\updefault}\special{ps: gsave 0 0 0 setrgbcolor}$\mbox{\bf T}_{\mbox{\scriptsize \bf 2}}$\special{ps: grestore}}}}
\put(6751,-61){\makebox(0,0)[lb]{\smash{\SetFigFont{11}{13.2}{\rmdefault}{\mddefault}{\updefault}\special{ps: gsave 0 0 0 setrgbcolor}$\mbox{\bf T}_{\mbox{\scriptsize \bf 3}}$\special{ps: grestore}}}}
\put(9151,-961){\makebox(0,0)[lb]{\smash{\SetFigFont{9}{10.8}{\rmdefault}{\mddefault}{\updefault}\special{ps: gsave 0 0 1 setrgbcolor}{\bf $a_1$}\special{ps: grestore}}}}
\put(9151,-2461){\makebox(0,0)[lb]{\smash{\SetFigFont{9}{10.8}{\rmdefault}{\mddefault}{\updefault}\special{ps: gsave 1 0 1 setrgbcolor}{\bf $a_2$}\special{ps: grestore}}}}
\put(9226,-4036){\makebox(0,0)[lb]{\smash{\SetFigFont{9}{10.8}{\rmdefault}{\mddefault}{\updefault}\special{ps: gsave 0 0 0 setrgbcolor}{\bf $a_3$}\special{ps: grestore}}}}
\put(9301,-5536){\makebox(0,0)[lb]{\smash{\SetFigFont{9}{10.8}{\rmdefault}{\mddefault}{\updefault}\special{ps: gsave 0 1 0 setrgbcolor}{\bf $a_4$}\special{ps: grestore}}}}
\put(9301,-7036){\makebox(0,0)[lb]{\smash{\SetFigFont{9}{10.8}{\rmdefault}{\mddefault}{\updefault}\special{ps: gsave 1 .88 .88 setrgbcolor}{\bf $a_5$}\special{ps: grestore}}}}
\put(9226,-8536){\makebox(0,0)[lb]{\smash{\SetFigFont{9}{10.8}{\rmdefault}{\mddefault}{\updefault}\special{ps: gsave .75 .38 0 setrgbcolor}{\bf $a_6$}\special{ps: grestore}}}}
\put(5476,-661){\makebox(0,0)[lb]{\smash{\SetFigFont{9}{10.8}{\rmdefault}{\mddefault}{\updefault}\special{ps: gsave 0 0 0 setrgbcolor}$\vartheta_1$\special{ps: grestore}}}}
\put(8701,-1336){\makebox(0,0)[lb]{\smash{\SetFigFont{9}{10.8}{\rmdefault}{\mddefault}{\updefault}\special{ps: gsave 0 0 0 setrgbcolor}$-\vartheta_1$\special{ps: grestore}}}}
\put(2251,-2461){\makebox(0,0)[lb]{\smash{\SetFigFont{9}{10.8}{\rmdefault}{\mddefault}{\updefault}\special{ps: gsave 0 0 0 setrgbcolor}$\vartheta_2$\special{ps: grestore}}}}
\put(8551,-3061){\makebox(0,0)[lb]{\smash{\SetFigFont{9}{10.8}{\rmdefault}{\mddefault}{\updefault}\special{ps: gsave 0 0 0 setrgbcolor}$-\vartheta_2$\special{ps: grestore}}}}
\put(2101,-3736){\makebox(0,0)[lb]{\smash{\SetFigFont{9}{10.8}{\rmdefault}{\mddefault}{\updefault}\special{ps: gsave 0 0 0 setrgbcolor}$\vartheta_3$\special{ps: grestore}}}}
\put(5251,-4636){\makebox(0,0)[lb]{\smash{\SetFigFont{9}{10.8}{\rmdefault}{\mddefault}{\updefault}\special{ps: gsave 0 0 0 setrgbcolor}$-\vartheta_3$\special{ps: grestore}}}}
\put(5401,-5461){\makebox(0,0)[lb]{\smash{\SetFigFont{9}{10.8}{\rmdefault}{\mddefault}{\updefault}\special{ps: gsave 0 0 0 setrgbcolor}$\vartheta_4$\special{ps: grestore}}}}
\put(2101,-5461){\makebox(0,0)[lb]{\smash{\SetFigFont{9}{10.8}{\rmdefault}{\mddefault}{\updefault}\special{ps: gsave 0 0 0 setrgbcolor}$\frac{\pi}{4}$\special{ps: grestore}}}}
\put(8476,-6061){\makebox(0,0)[lb]{\smash{\SetFigFont{9}{10.8}{\rmdefault}{\mddefault}{\updefault}\special{ps: gsave 0 0 0 setrgbcolor}$-\vartheta_4-\frac{\pi}{4}$\special{ps: grestore}}}}
\put(2251,-7111){\makebox(0,0)[lb]{\smash{\SetFigFont{9}{10.8}{\rmdefault}{\mddefault}{\updefault}\special{ps: gsave 0 0 0 setrgbcolor}$\vartheta_5$\special{ps: grestore}}}}
\put(5551,-7036){\makebox(0,0)[lb]{\smash{\SetFigFont{9}{10.8}{\rmdefault}{\mddefault}{\updefault}\special{ps: gsave 0 0 0 setrgbcolor}$\frac{\pi}{4}$\special{ps: grestore}}}}
\put(8026,-7636){\makebox(0,0)[lb]{\smash{\SetFigFont{9}{10.8}{\rmdefault}{\mddefault}{\updefault}\special{ps: gsave 0 0 0 setrgbcolor}$-\vartheta_5-\frac{\pi}{4}$\special{ps: grestore}}}}
\put(2251,-8611){\makebox(0,0)[lb]{\smash{\SetFigFont{9}{10.8}{\rmdefault}{\mddefault}{\updefault}\special{ps: gsave 0 0 0 setrgbcolor}$\vartheta_6$\special{ps: grestore}}}}
\put(4576,-9811){\makebox(0,0)[lb]{\smash{\SetFigFont{9}{10.8}{\rmdefault}{\mddefault}{\updefault}\special{ps: gsave 0 0 0 setrgbcolor}$-\vartheta_6-\frac{\pi}{2}$\special{ps: grestore}}}}
\put(8701,-8911){\makebox(0,0)[lb]{\smash{\SetFigFont{9}{10.8}{\rmdefault}{\mddefault}{\updefault}\special{ps: gsave 0 0 0 setrgbcolor}$\frac{\pi}{2}$\special{ps: grestore}}}}
\put(1351,-10111){\makebox(0,0)[lb]{\smash{\SetFigFont{10}{12.0}{\rmdefault}{\mddefault}{\updefault}\special{ps: gsave 0 0 0 setrgbcolor}$x^4$\special{ps: grestore}}}}
\put( 76,-8986){\makebox(0,0)[lb]{\smash{\SetFigFont{10}{12.0}{\rmdefault}{\mddefault}{\updefault}\special{ps: gsave 0 0 0 setrgbcolor}$x^5$\special{ps: grestore}}}}
\put(4576,-10111){\makebox(0,0)[lb]{\smash{\SetFigFont{10}{12.0}{\rmdefault}{\mddefault}{\updefault}\special{ps: gsave 0 0 0 setrgbcolor}$x^6$\special{ps: grestore}}}}
\put(3301,-8986){\makebox(0,0)[lb]{\smash{\SetFigFont{10}{12.0}{\rmdefault}{\mddefault}{\updefault}\special{ps: gsave 0 0 0 setrgbcolor}$x^7$\special{ps: grestore}}}}
\put(8551,-10111){\makebox(0,0)[lb]{\smash{\SetFigFont{10}{12.0}{\rmdefault}{\mddefault}{\updefault}\special{ps: gsave 0 0 0 setrgbcolor}$x^8$\special{ps: grestore}}}}
\end{picture}